\documentclass[aps,twocolumn,a4]{revtex4}
\usepackage{amsmath}
\usepackage{latexsym}
\usepackage{float}
\usepackage{amssymb}
\usepackage{graphicx}
\usepackage{epsfig}

\begin{document}

\title{Structure of bottle-brush polymers in solution: 
A Monte Carlo test of models for the scattering function}
\author{Hsiao-Ping Hsu, Wolfgang Paul, and Kurt Binder}
\affiliation{Institut f\"ur Physik, Johannes Gutenberg-Universit\"at Mainz\\
D-55099 Mainz, Staudinger Weg 7, Germany}

\date{\today}
\begin{abstract}
Extensive Monte Carlo results are presented for the structure of
a bottle-brush polymer under good solvent or theta solvent
conditions. Varying the side chain length, backbone length, and
the grafting density for a rigid straight backbone, both radial
density profiles of monomers and side chain ends are obtained, as
well as structure factors describing the scattering from a single
side chain and from the total bottle-brush polymer. To describe
the structure in the interior of a very long bottle-brush, a
periodic boundary condition in the direction along the backbone is
used, and to describe effects due to the finiteness of the
backbone length, a second set of simulations with free ends of the
backbone is performed. In the latter case, the inhomogeneity of
the structure in the direction along the backbone is carefully
investigated. We use these results to test various
phenomenological models that have been proposed to interpret
experimental scattering data for bottle-brush macromolecules.
These models aim to extract information on the radial density
profile of a bottle-brush from the total scattering via suitable
convolution approximations. Limitations of this approach and the
optimal way to perform the analysis of the scattering data within this
approach are discussed.
\end{abstract}

\maketitle

\section{Introduction}

Recently there has been a great experimental 
(see e.~g.~\cite{1,1a,2,3,4,4a,4b}) 
and theoretical 
~\cite{5,6,7,8,9,10,11,12,13,14,15,16,17,18,19,20,21,22,23,
24,25,26,27,28,29,30,31,32} interest in the
conformation of so-called bottle-brush polymers. Such polymers
consist of a long (flexible) main chain, at which many flexible
(shorter) side chains are densely grafted, such that an overall
shape of a worm-like cylindrical brush results~\cite{1,1a,2,3,4}.
Synthesizing such polymers with suitable characteristics, materials
can be prepared whose properties can be adjusted by external
stimuli, such as the solvent quality, pH, or temperature, and this
fact makes such bottle-brush polymers interesting for various
applications~\cite{33,34}. For controlling the properties of such
bottle-brush polymers, a good theoretical understanding of their
structure and conformation, as a function of control parameters
such as the chain lengths of the main chain and the side chains,
their grafting density, and the solvent quality, is mandatory.
However, despite the longstanding theoretical 
efforts~\cite{5,6,7,8,9,10,11,12,13,14,15,16,17,18,19,20,21,22,23,
24,25,26,27,28,29,30,31,32}
this problem is still incompletely understood. While one has
various scaling predictions (see~\cite{31} for a recent review)
and treatments based on the strong stretching limit of
self-consistent field theory exist since a long time 
(e.g. Refs.~\cite{5,6,7,8}), recent simulations~\cite{31} and also the
experiments~\cite{2,3,4} indicate that the regime where these
theories become accurate would require side chain lengths that are
hardly accessible either by experiment or by simulation. As a
consequence, the theoretical guidance for the interpretation of
extensive experiments by combined light and small-angle neutron
scattering analysis~\cite{2,3,4} is still incomplete.

In the present work, we make a contribution to clarify this
problem by extensive Monte Carlo simulations of bottle-brush
polymers~\cite{31,32} using the ''Pruned-Enriched Rosenbluth
Method'' (PERM algorithm)~\cite{35} to obtain the relevant
information on the conformation of bottle-brush polymers under
various conditions, that are needed to test the phenomenological
models used to interpret the experimental scattering 
data~\cite{2,3,4}. As was shown in~\cite{31} the PERM algorithm is very
powerful to obtain a wealth of simulation data for the case of
side chains grafted to a hard rod (a generalization of the algorithm
to a flexible backbone is far from trivial), representing a strictly rigid
backbone polymer. This idealization describes a real bottle-brush
chain only
locally. However, all theoretical models used for the analysis of
experiments~\cite{2,3,4} determining the structure of a bottle-brush
do contain the rigid backbone as a special case. It is this case
for which we can undertake a stringent 
test of physical model assumptions underlying the analysis of
experimental data. Of course, there is no reason to assume that a model
that already fails for the (simpler) rigid backbone case should be
accurate for bottle-brushes with flexible backbones.

In Sec.~II, we introduce our model and recall the most basic facts
on our simulation method and define the quantities that are
studied. In Sec.~III, we give a comprehensive overview of our
results on various physical properties of the bottle-brush
polymers. Sec.~IV then is devoted to the problem relevant for the
interpretation of the experiments, namely the test of theoretical
models used in~\cite{2,3,4,4a} for our system: Note that unlike the
experiment, we can extract information of radial density profiles
and geometrical characteristics of individual side chains directly
from the simulation, simultaneously with but independent of the
information gathered on the scattering functions, and thus a
stringent test of the proposed relations between these quantities
is possible. Sec.~V then summarizes our conclusions.

\section{MODEL AND SIMULATION METHODOLOGY}

As in Ref.~\cite{31}, we use as a coarse-grained simple model of
flexible polymers in solution, the self-avoiding walk on the simple
cubic lattice. Each lattice site can be occupied by a single
effective monomeric unit only, and this excluded volume
interaction corresponds to polymer chain conformations under good
solvent conditions~\cite{36,37,38}. Introducing an energy
parameter $\epsilon$ that is won if two effective monomers occupy
nearest neighbor sites, one can describe variable solvent quality
in this model simply by varying the temperature $T$: the Theta
temperature $\Theta$ where this attraction approximately cancels
the excluded volume repulsion, in the sense that the mean square
gyration radius $\langle R_g^2\rangle _T$ of a chain scales
linearly with the chain length $N$, apart from logarithmic
corrections \cite{36}, occurs for $q_\Theta =\exp
(-\epsilon/k_B\Theta)=1.3087$~\cite{35}. We shall present results
both for $T=\Theta$ and for $T \rightarrow \infty$ \{where
$q=\exp(-\epsilon/k_BT)=1$, and hence only the excluded volume
interaction is present\}, in view of the fact that most cases of
experimental interest will be somewhere in between these limits.

Following Ref.~\cite{31} we take the rigid backbone along the $z$-axis
of our coordinate system. Using even values for the length
$L_b$ of the backbone, measuring all lengths in units of the
lattice spacings, grafting sites $z=1$, $z=L_b$ at the end of the
backbone are labeled as $s_1$, sites adjacent to the ends
($z=2$, $z=L_{b-1})$ as $s_2$, and so on, until the center of the
bottle-brush, sites $z=L_b/2$, $z=L_b/2+1$ being denoted as
$s_{L_b/2}$. Of course, a dependence of properties of a side chain
on the coordinate $s_k$ occurs only when we consider bottle-brush
polymers with free ends in the $z$-direction, while no dependence on
$s_k$ occurs if we choose periodic boundary conditions (pbcs) in
$z$-direction such that the grafting site $z=L_b$ is nearest
neighbor of $z=1$: in this case full translational invariance in
$z$-direction holds, and the distribution function of the monomers
$P_{z'}(z-z')$ of a side chain grafted at $z'$ must be symmetric
around $z'$, $P_{z'}(z-z')=P_{z'}(z'-z)$. This symmetry property
does not hold only for the distribution function of all the
monomers that belong to that side chain, but also for individual
monomers $i=1,\ldots,N$ along the side chain, in particular for
the chain ends. Also the average $z$-coordinate of the center of
mass of the side chain coincides with $z'$. None of these
symmetries carry over to the case with free ends, of course; in
the latter case the monomers of the side chain can have $z$-coordinates 
in the region $-N+1\leq z \leq L_b+N$ (the boundaries
of this interval do not occur in practice, of course, it would
require that a side chain grafted at $z=1$ or $z=L_b$ is linearly
stretched out in the $-z$ or $+z$-direction, respectively).

When one considers properties of individual side chains, which are
stretched away from the backbone, two non-equivalent directions
$x$, $y$ need to be distinguished~\cite{31}: 
defining the vector toward to the center of mass (C.M.) of a chain from
its grafting point $z'$ as $(X_{z'},Y_{z'},Z_{z'})$  in
a fixed laboratory frame, for a particular configuration of the
side chain, we can define the $x$-axis along the vector $(X_{z'},Y_{z'})$,
and require that the $y$-axis
is perpendicular to the $x$-axis and also lies in the
$X_{z'}-Y_{z'}$ plane. Since for a densely grafted bottle-brush
polymer strong stretching of the side chains is 
expected~\cite{5,6,7,8,9,10,16} this distinction allows to compute linear
dimensions of the side chains in the direction along which the
chain is stretched, and perpendicular to it.

In practice, side chain lengths up to $N=50$ were considered,
while choices $L_b=32$, $64$, $128$ and 256 as well as two values of the
grafting density, $\sigma =1/2$ and $\sigma = 1$, were considered.
A distinctive feature of our implementation of the PERM algorithm
is~\cite{31} that in one run one gets information on properties
for all integer values of N from $N=1$ up to $N_{\rm max }$
\{which in our case was chosen to be $N_{\rm max }=50$, so the
largest polymer simulated had a total number of monomers
$\mathcal{N}_{\rm tot}=L_b+L_b \sigma N=256 +
(256/2)50=6656$, since for $L_b=256$ the case $\sigma =1$ no
longer was feasible\}. For details on the implementation of the
PERM algorithm for bottle-brush polymers, we refer the reader to
Ref.~\cite{31}.

\begin{figure}
\begin{center}
(a)\includegraphics[scale=0.29,angle=270]{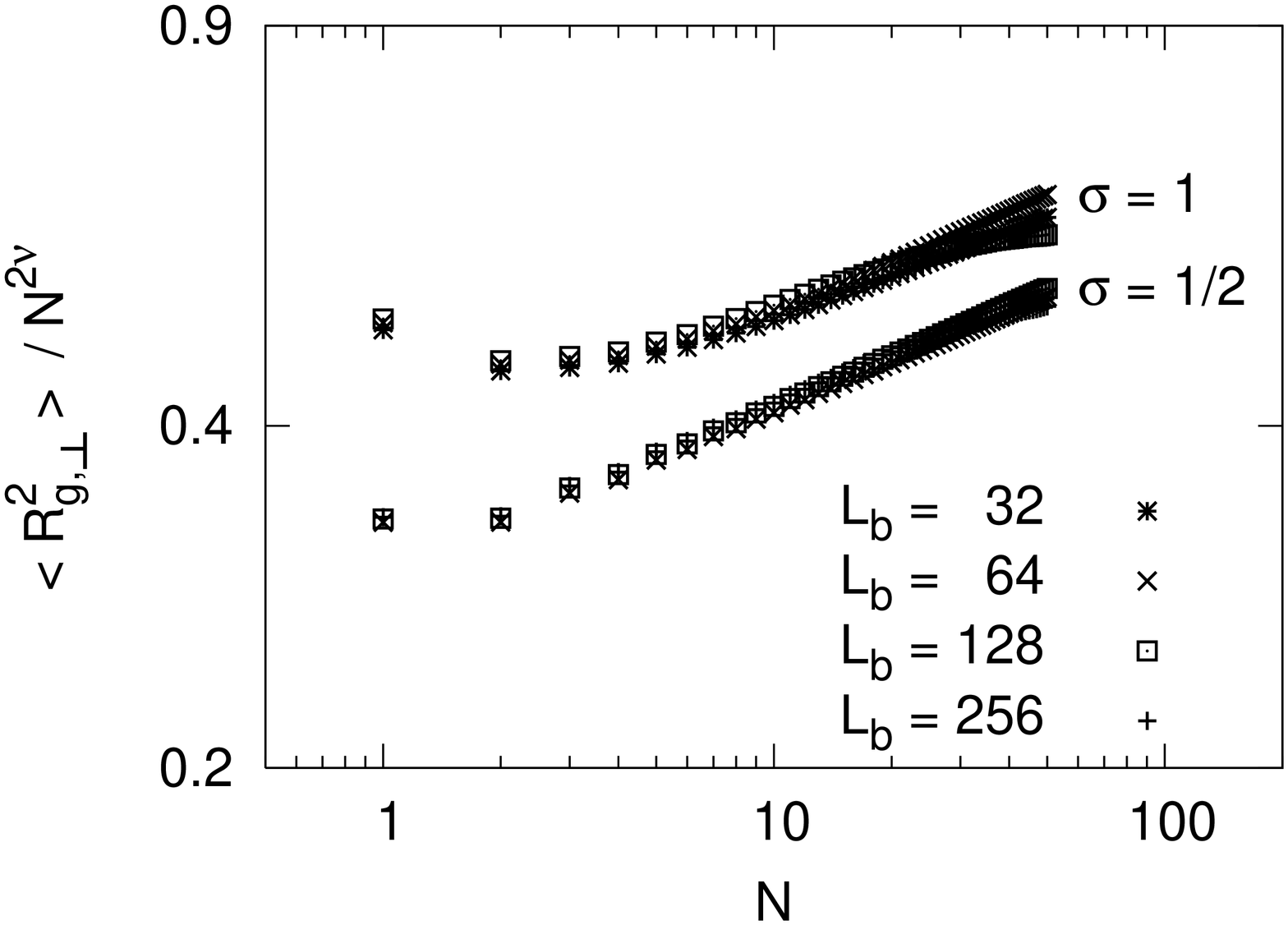} \hspace{0.4cm}
(b)\includegraphics[scale=0.29,angle=270]{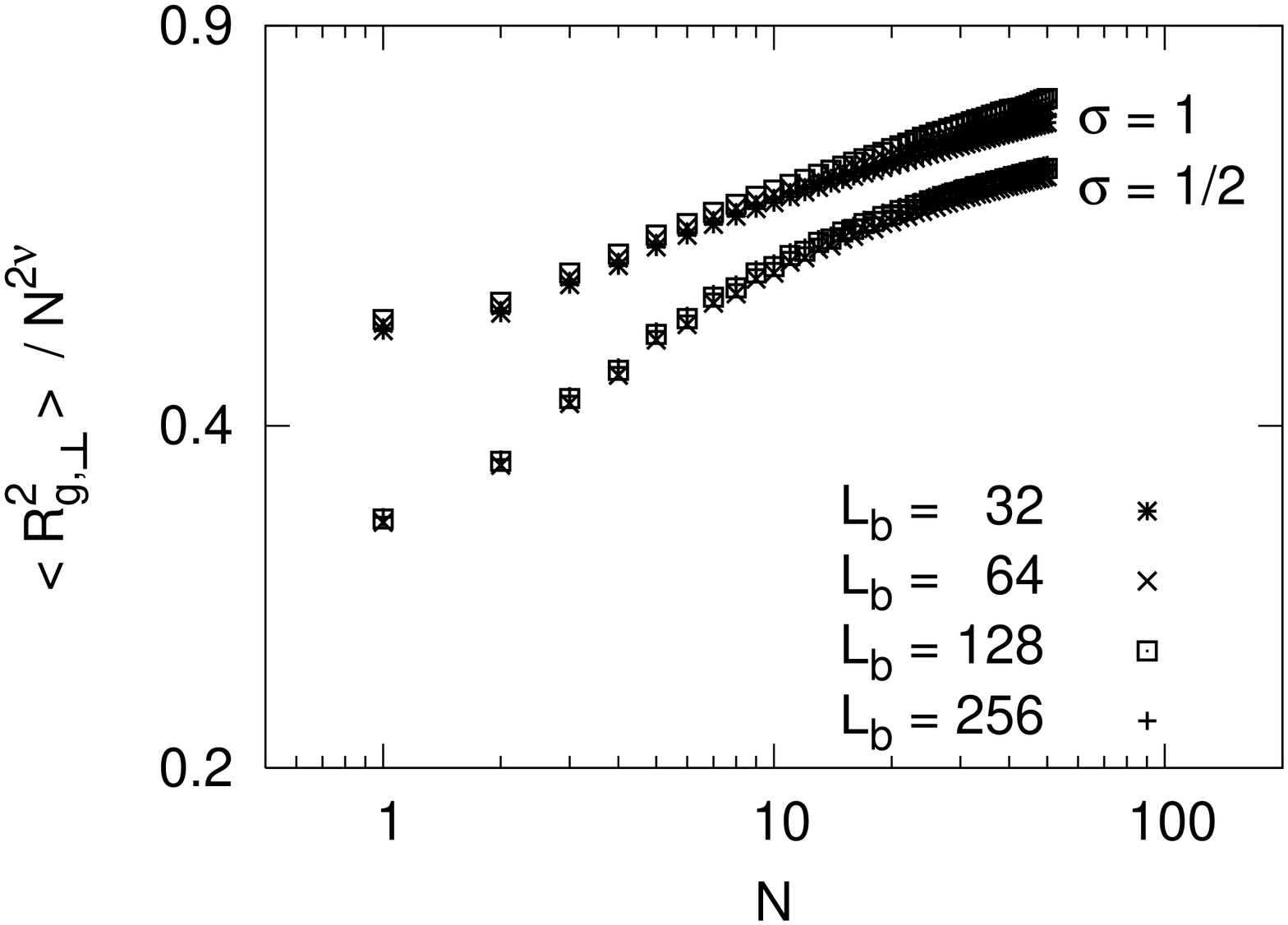}
\caption{Log-log plot of the rescaled
mean square gyration radius
perpendicular to the backbone, $\langle R_{g,\bot}^2\rangle
/N^{2\nu}$ of the whole bottle-brush versus the side chain length,
for a good solvent where $\nu = 0.588 $ (a) and a $\Theta$-solvent
where $\nu = 0.5$ was taken (b). Two choices of $\sigma $ and four
choices of $L_b$ are included, as indicated. All data are for
bottle-brushes with free ends.}
\label{fig1}
\end{center}
\end{figure}

\begin{figure*}
\begin{center}
(a)\includegraphics[scale=0.29,angle=270]{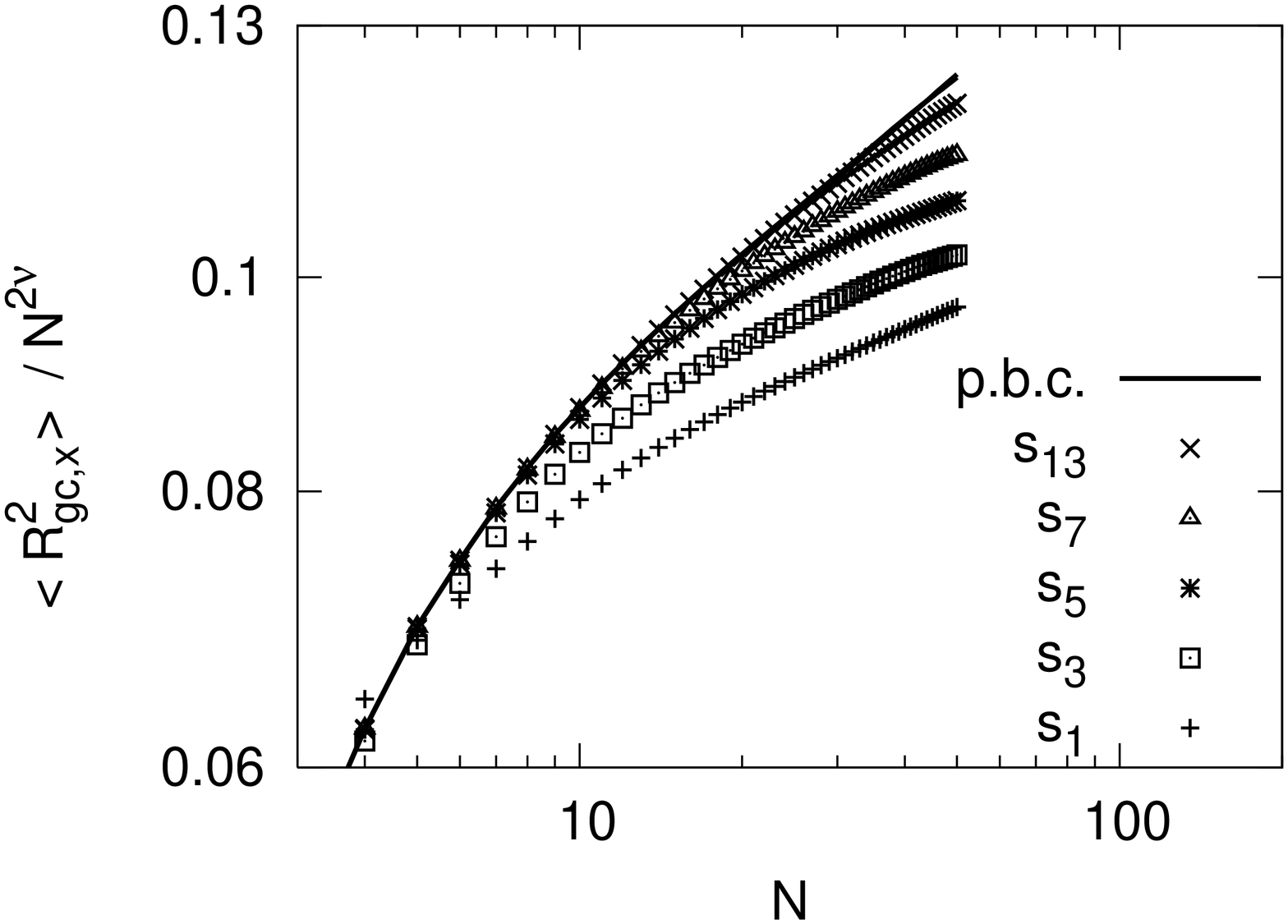}\\
(b)\includegraphics[scale=0.29,angle=270]{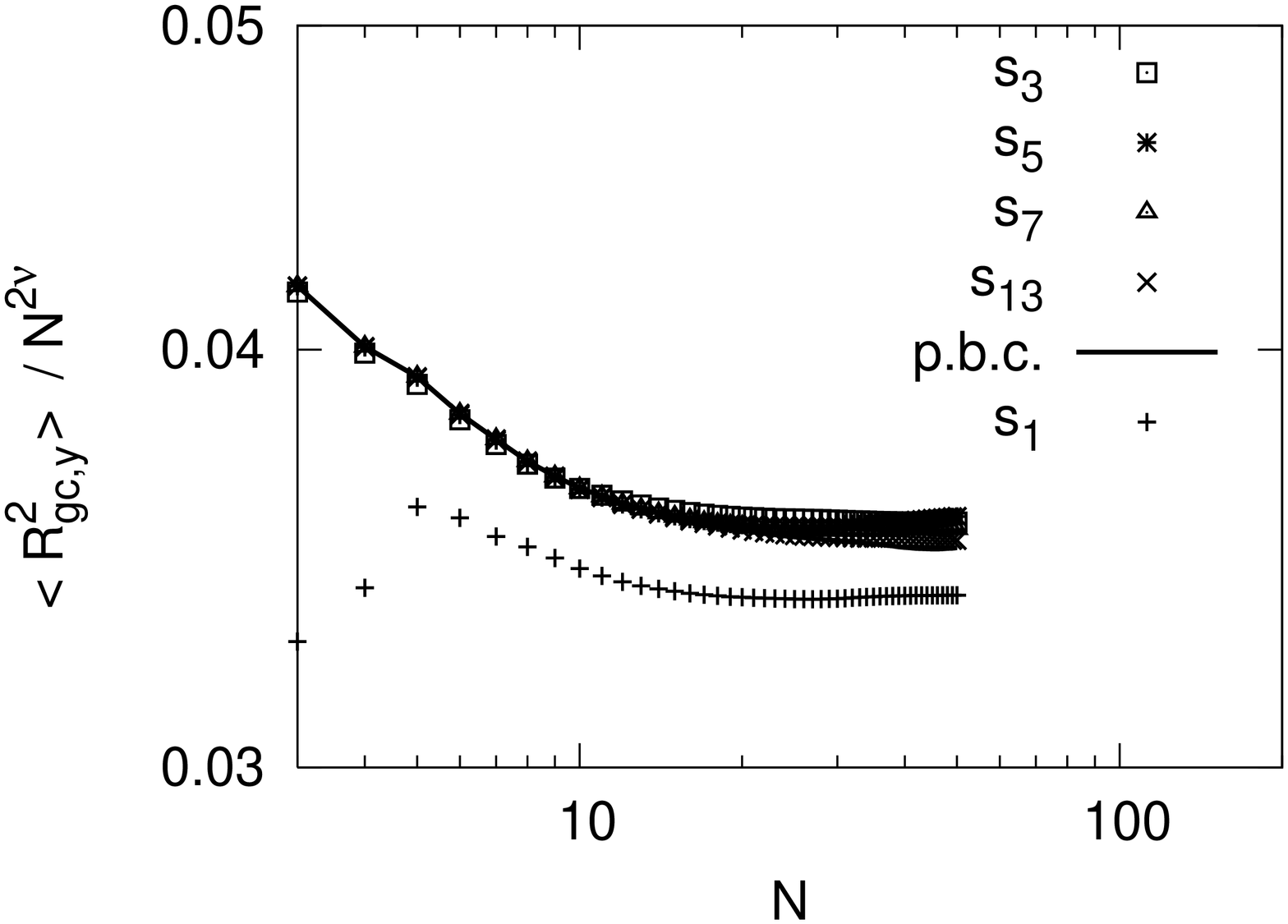}\\
(c)\includegraphics[scale=0.29,angle=270]{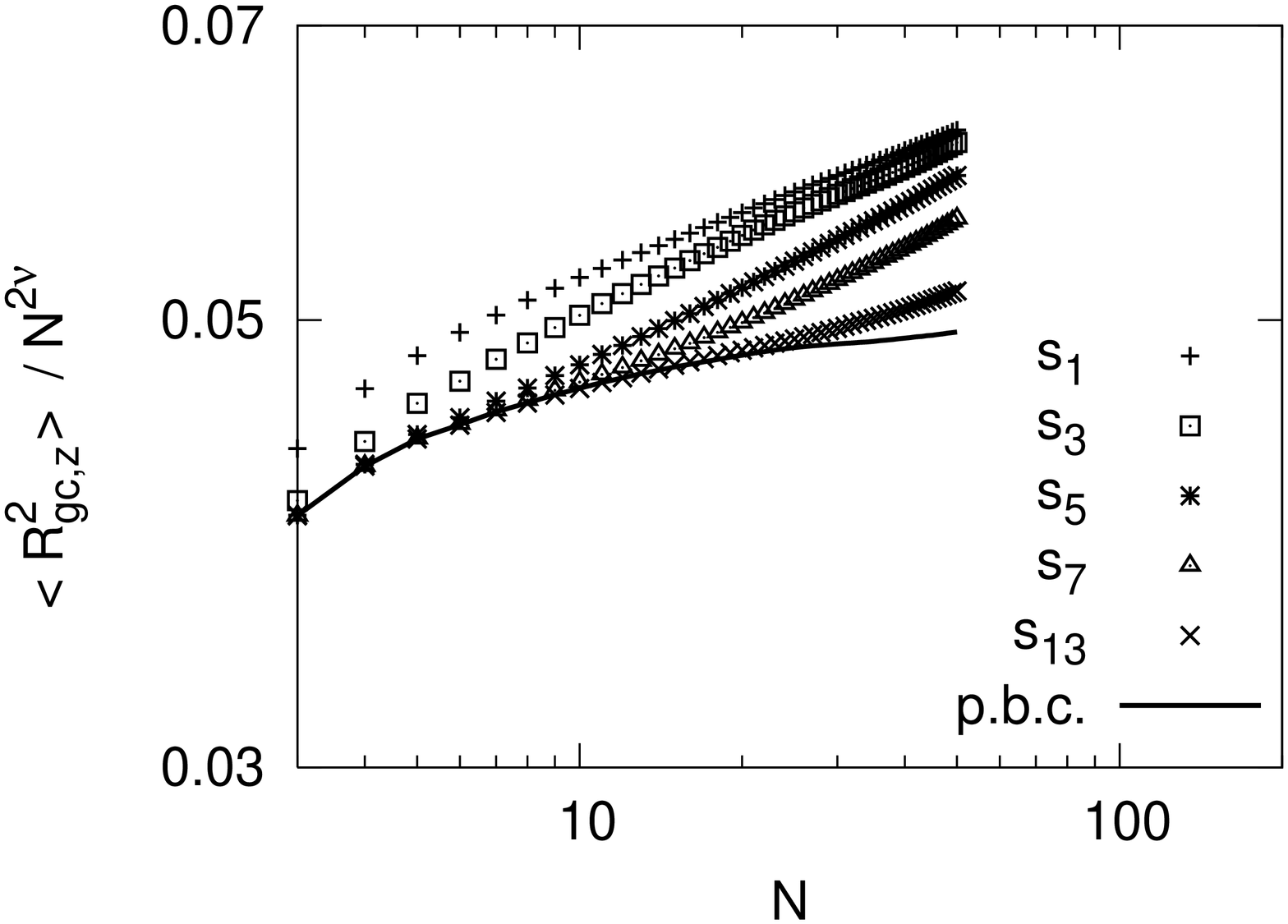}
\caption{Log-log plot of the rescaled
mean square gyration radii of the side chains, $\langle
R_{gc,x}^2 \rangle/N^{2\nu}$ (a), $\langle R_{gc,y}^2 \rangle
/N^{2\nu}$ (b) and $\langle R_{gc,z}^2\rangle/N^{2\nu}$ (c) versus
the side chain length N, for the good solvent case, $L_b=32,
\sigma =1$ and various choices of the grafting sites, as shown by
the coordinate $s_k$ (cf. Sec.~II for explanations). The full
curves show analogous data for the case of pbcs.}
\label{fig2}
\end{center}
\end{figure*}

\begin{figure*}
\begin{center}
(a)\includegraphics[scale=0.29,angle=270]{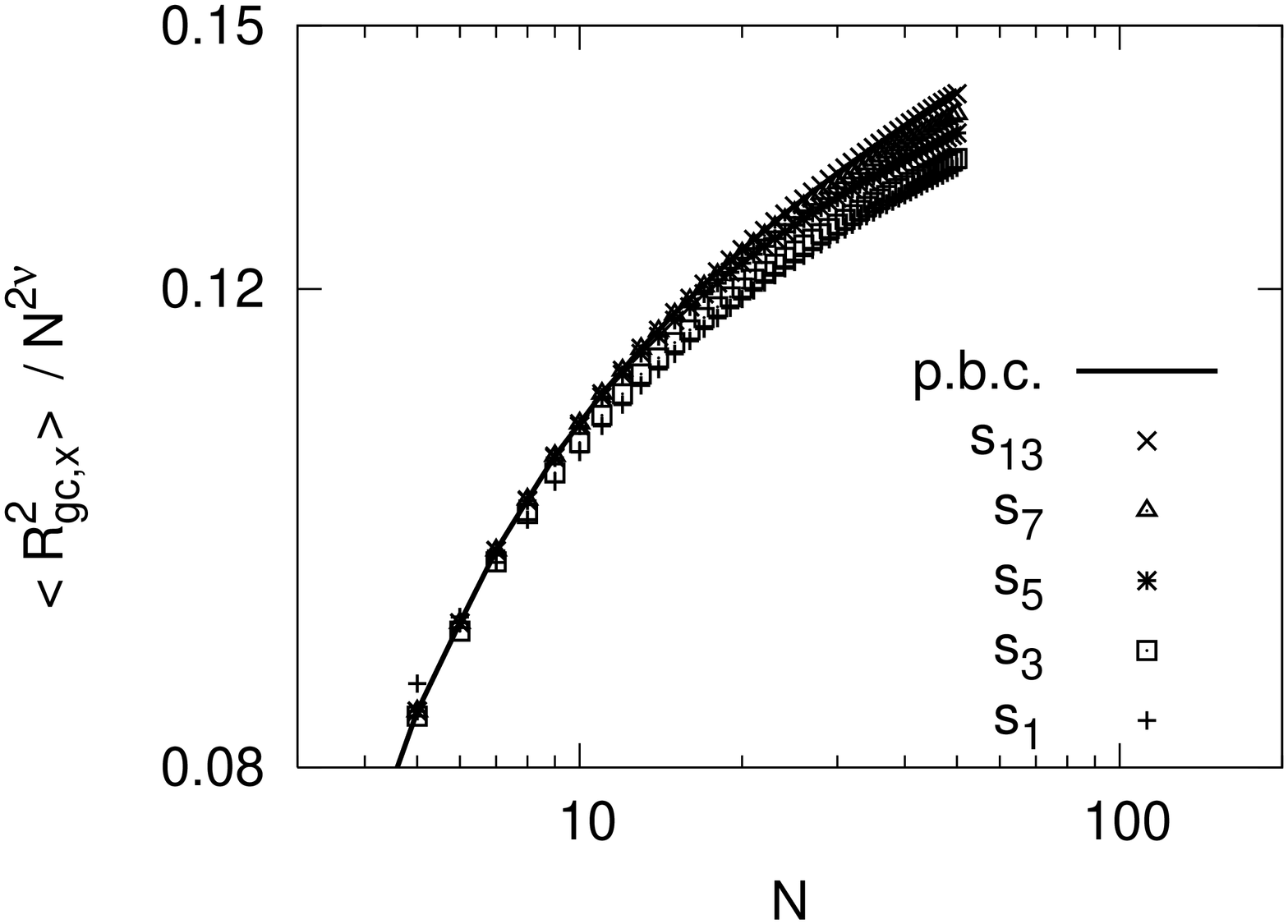}\\
(b)\includegraphics[scale=0.29,angle=270]{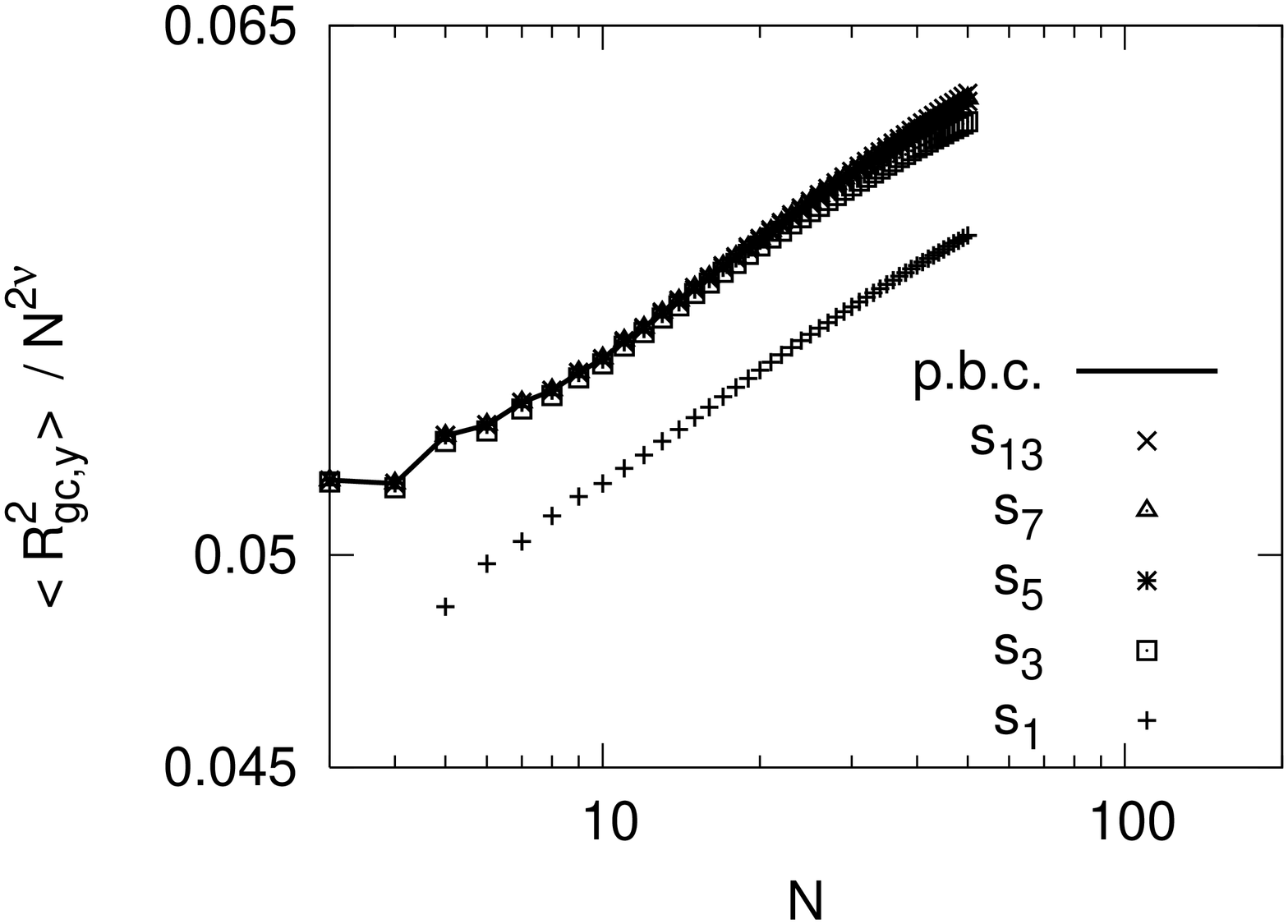}\\
(c)\includegraphics[scale=0.29,angle=270]{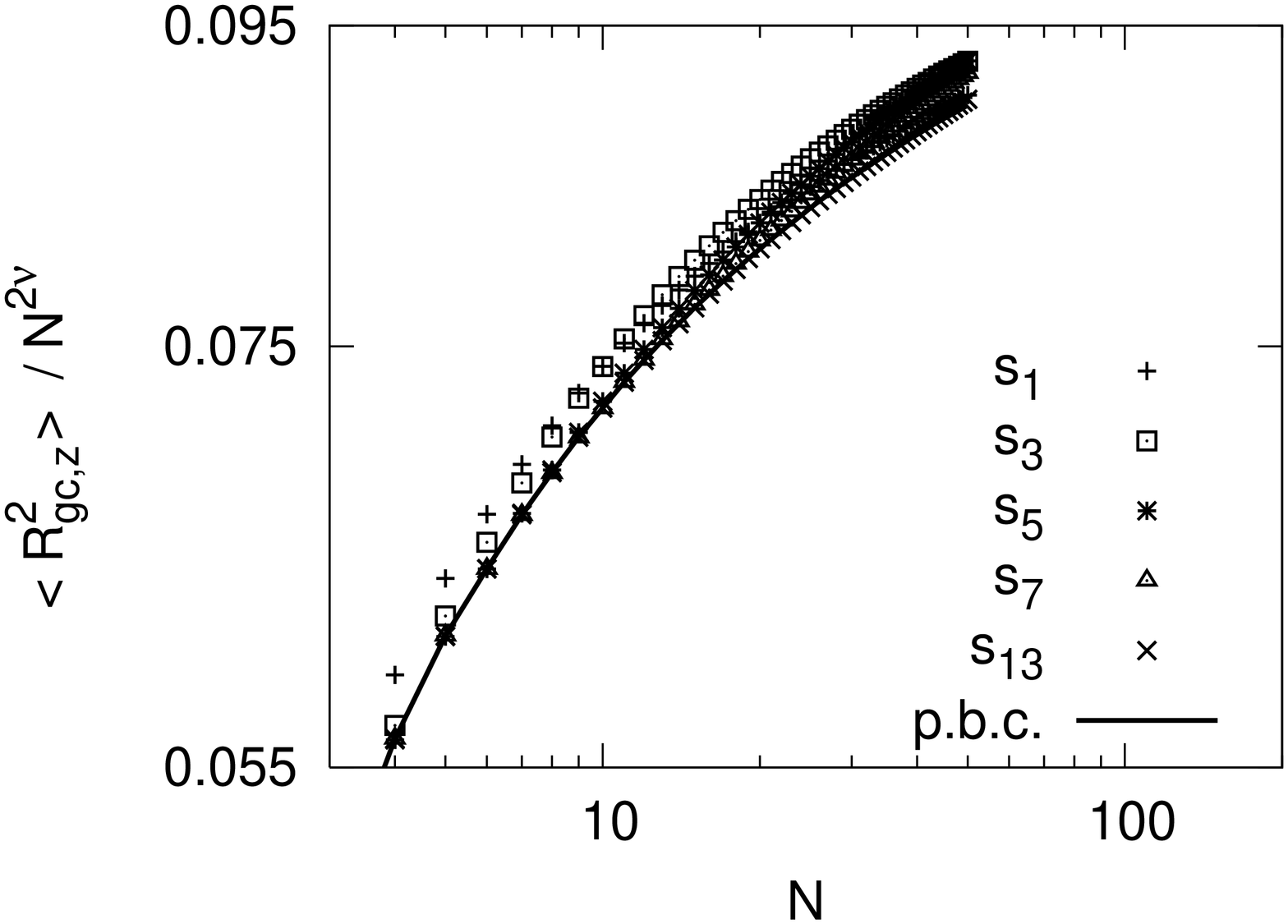}
\caption{Same as Fig.~\ref{fig2}, but
for the case of the $\Theta$-solvent.}
\label{fig3}
\end{center}
\end{figure*}

\begin{figure*}
\begin{center}
\vspace{1cm}
(a)\includegraphics[scale=0.29,angle=270]{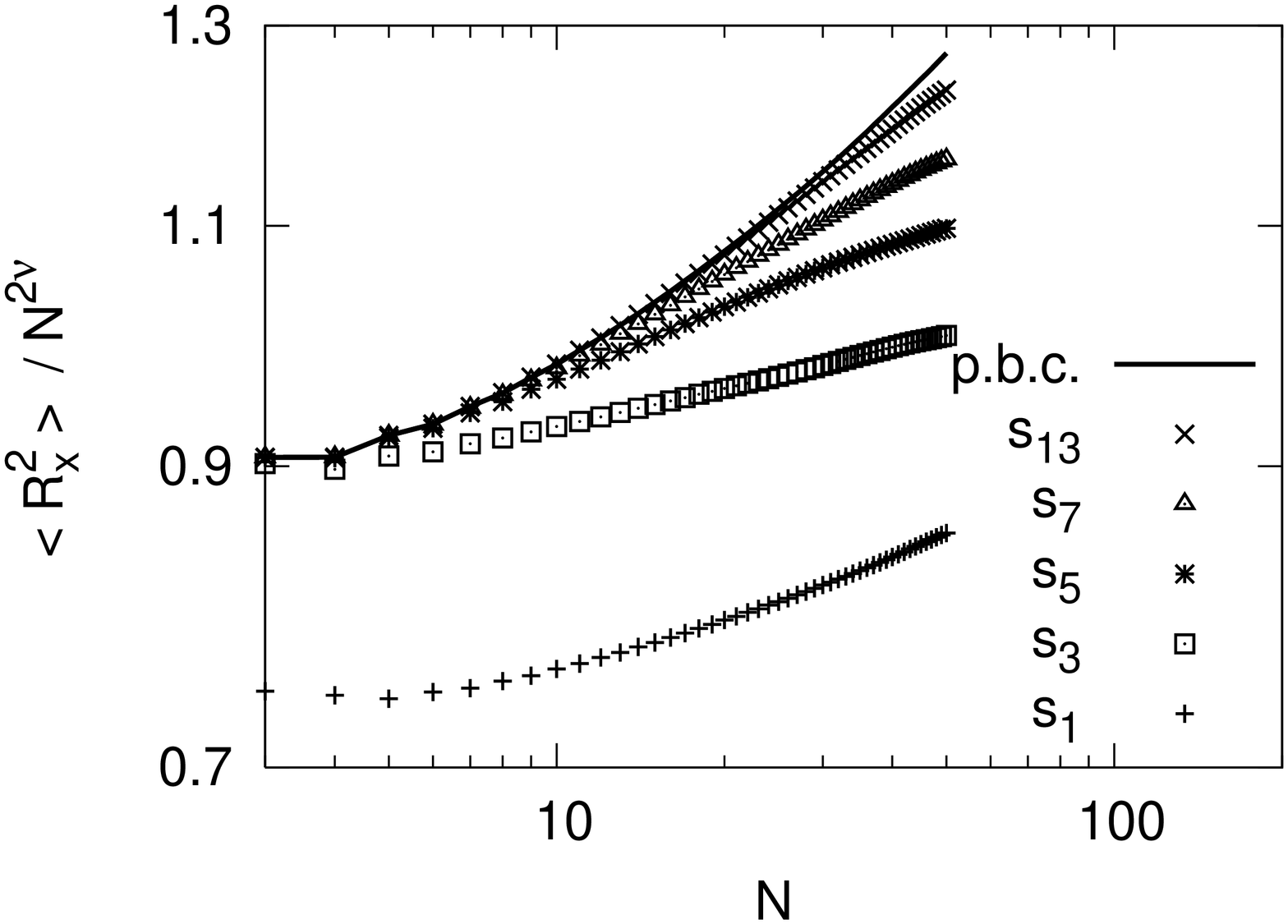}\hspace{0.4cm}
(b)\includegraphics[scale=0.29,angle=270]{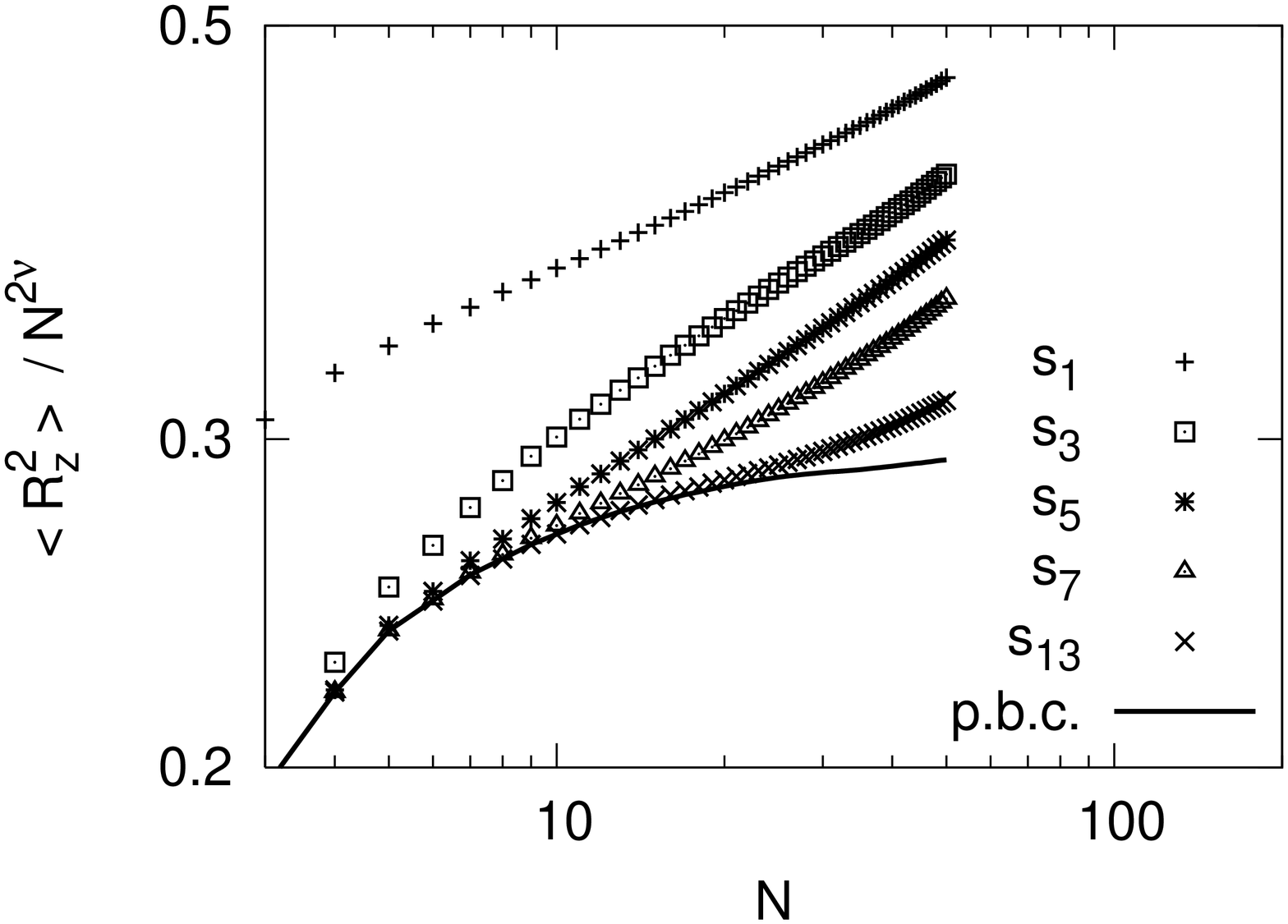}
(c)\includegraphics[scale=0.29,angle=270]{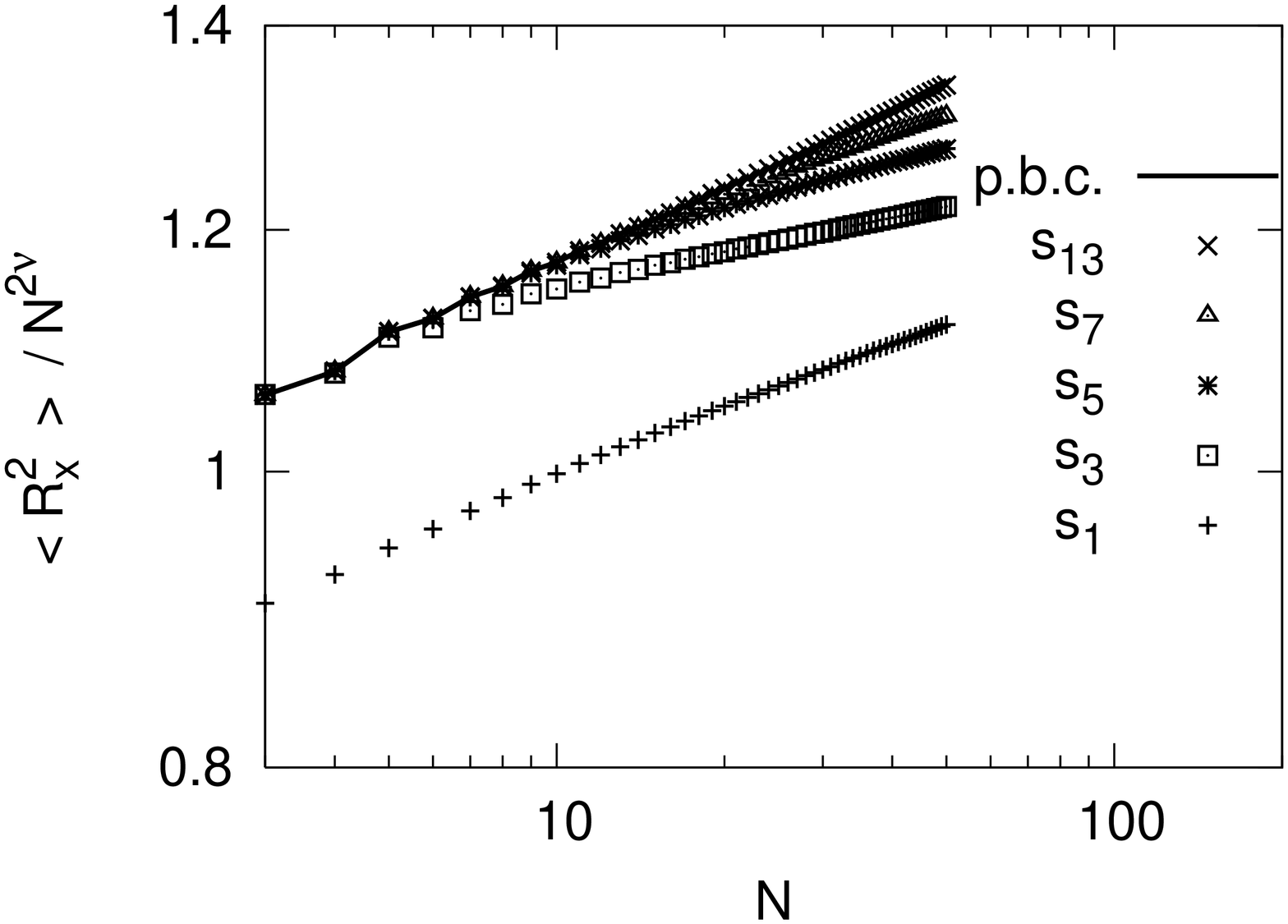}\hspace{0.4cm}
(d)\includegraphics[scale=0.29,angle=270]{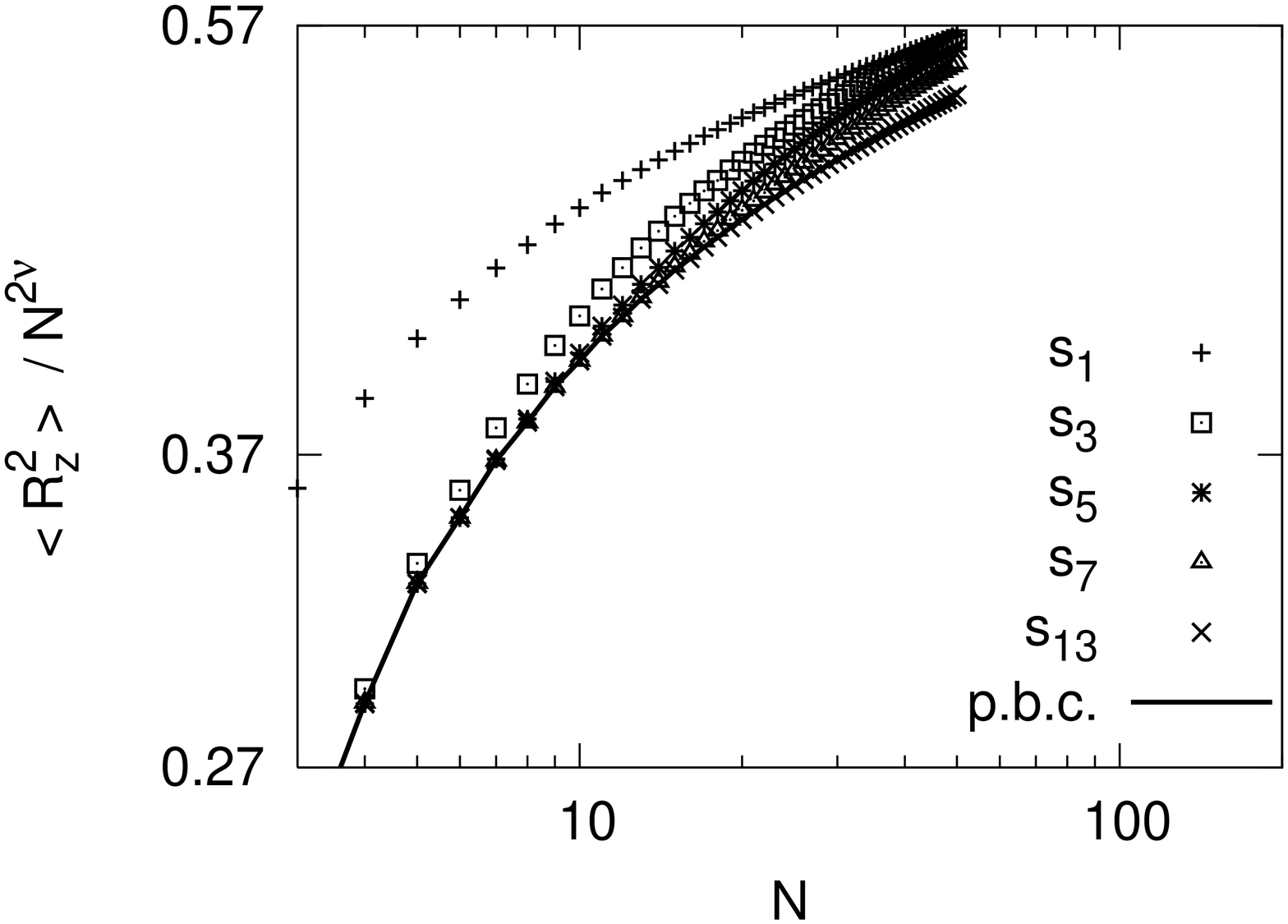}
\caption{Log-log plot of the mean square
backbone-to-end distance of the side chains versus the side chain
length $N$, for $L_b=32$ and $\sigma =1$. Panels (a,b) refer to
the good solvent, panels (c,d) to the $\Theta$-solvent case. The
components shown are $\langle R_x^2 \rangle /N^{2\nu}$ (a,c) and
$\langle R_z^2 \rangle /N^{2 \nu}$ (b,d). Symbols denote different
coordinates $s_k$ along the backbone, while the full curves show
the analogous result for pbcs.}
\label{fig4}
\end{center}
\end{figure*}

\begin{figure*}
\begin{center}
\vspace{1cm}
(a)\includegraphics[scale=0.29,angle=270]{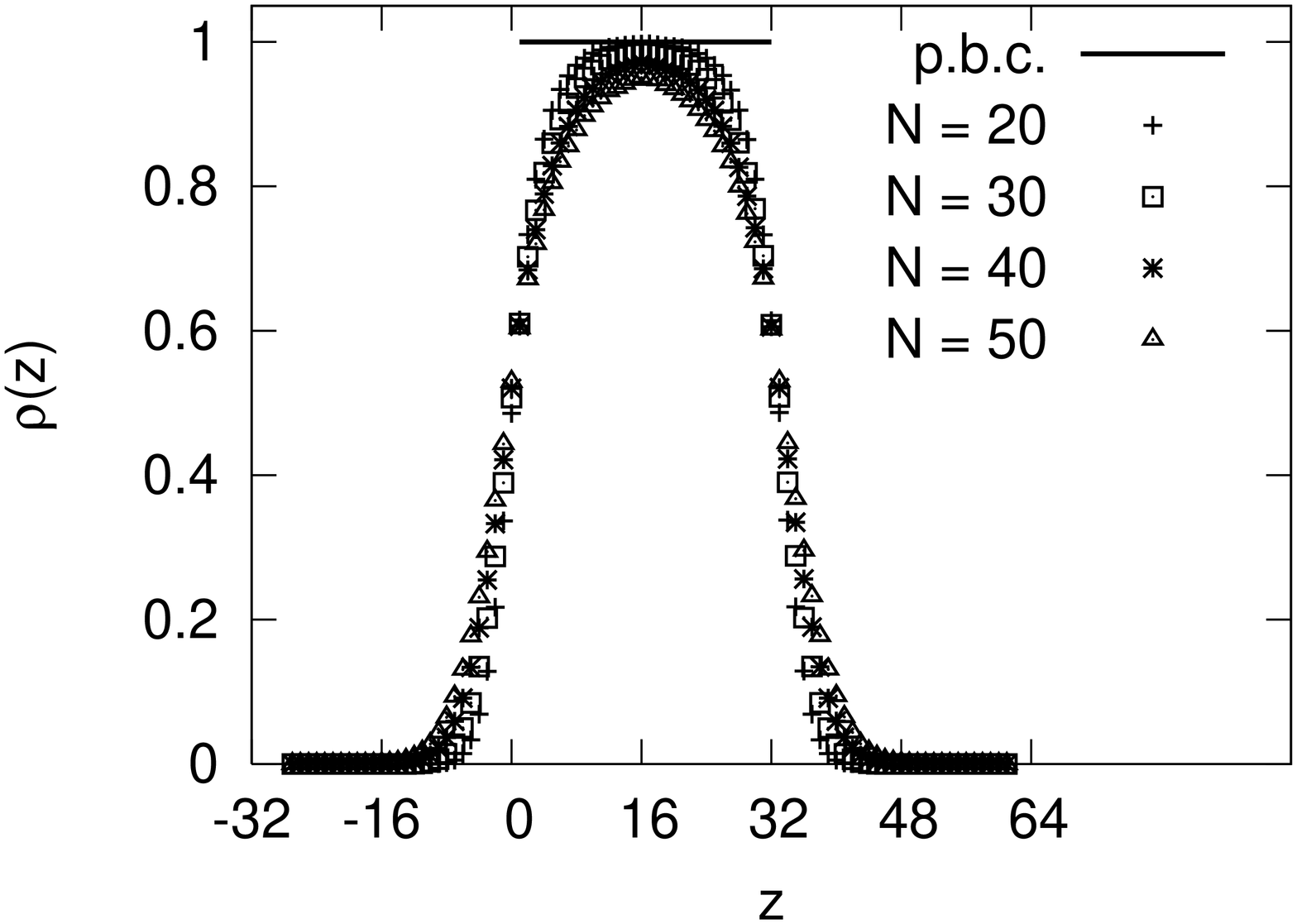}\hspace{0.4cm}
(b)\includegraphics[scale=0.29,angle=270]{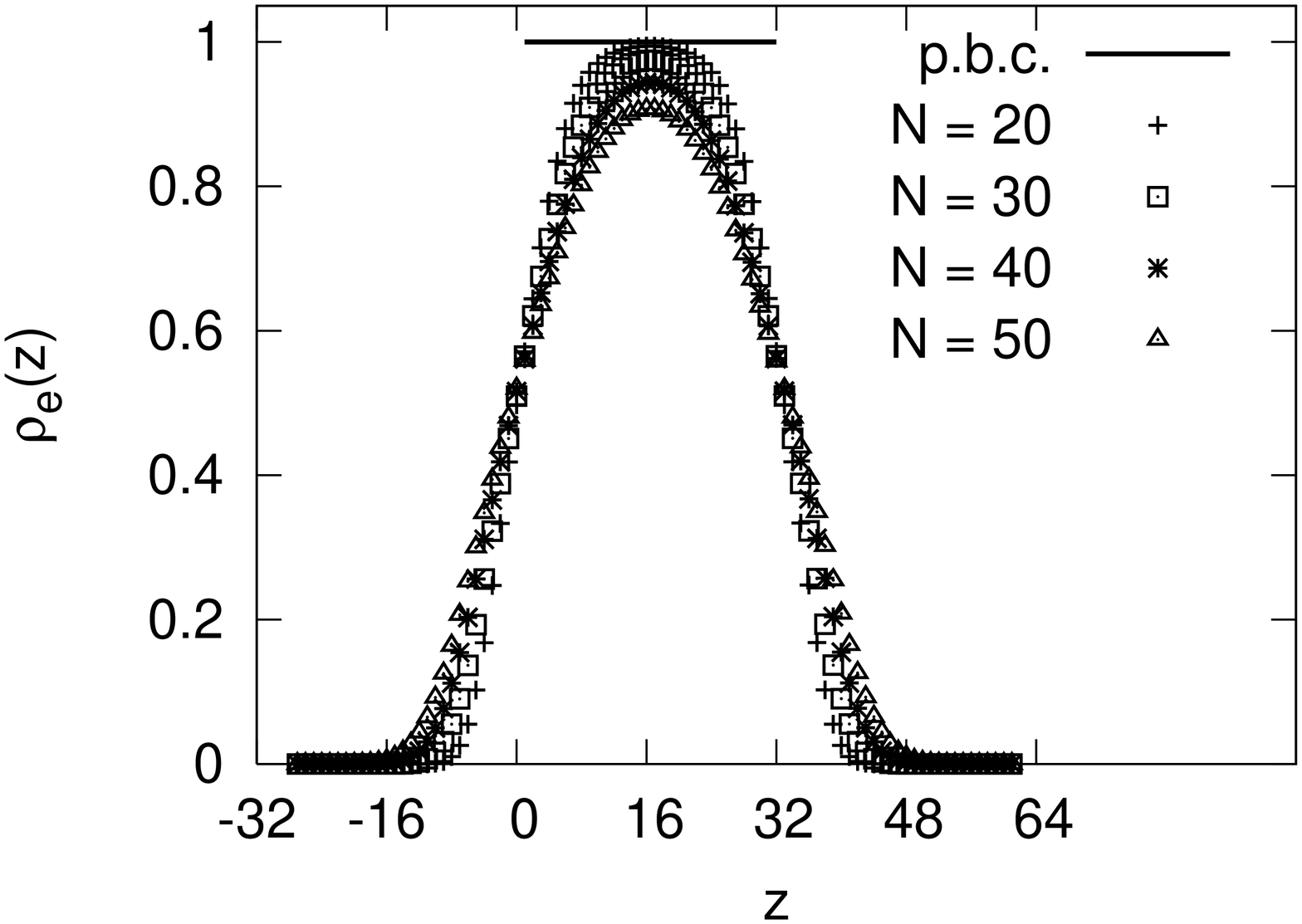}
(c)\includegraphics[scale=0.29,angle=270]{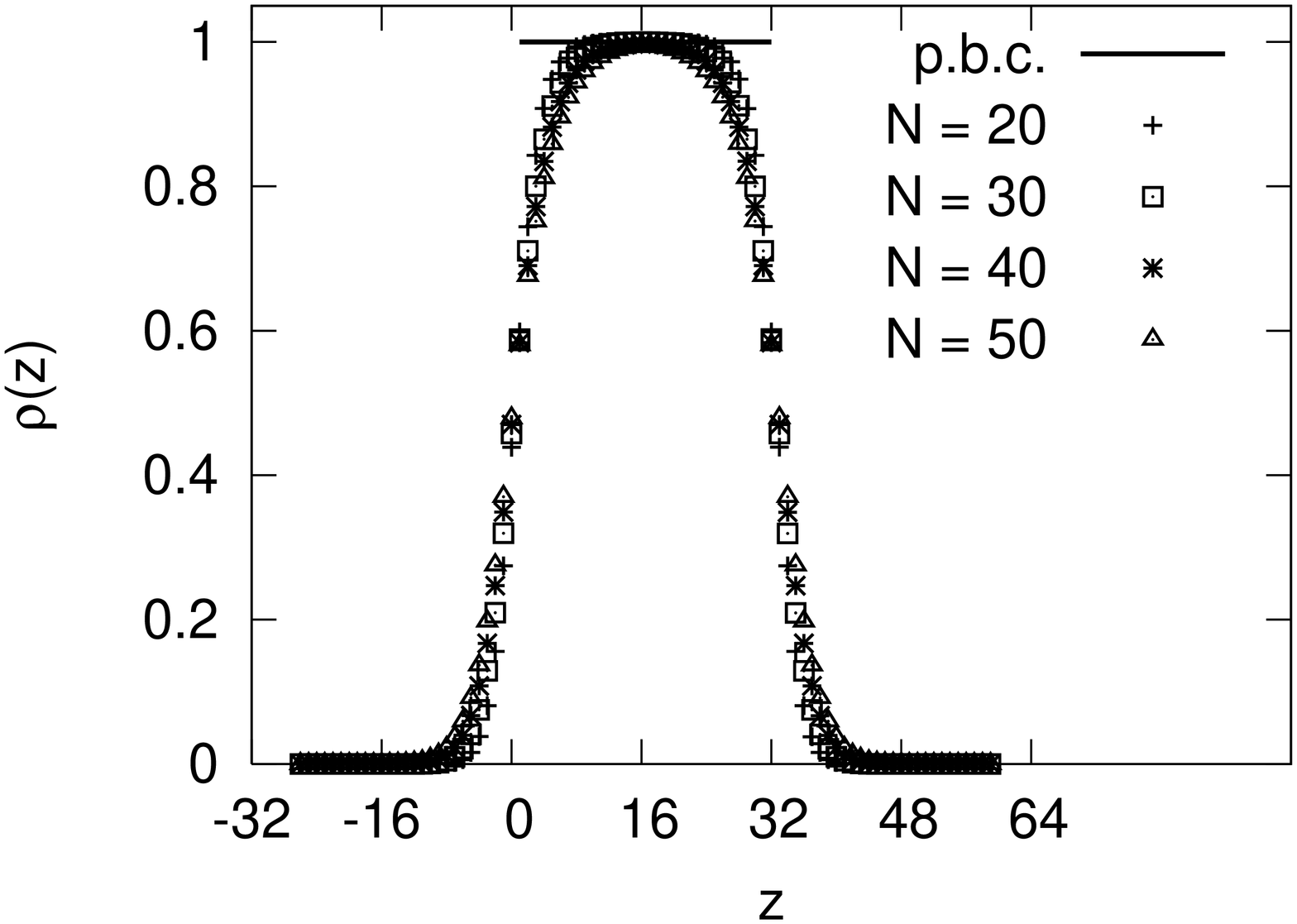}\hspace{0.4cm}
(d)\includegraphics[scale=0.29,angle=270]{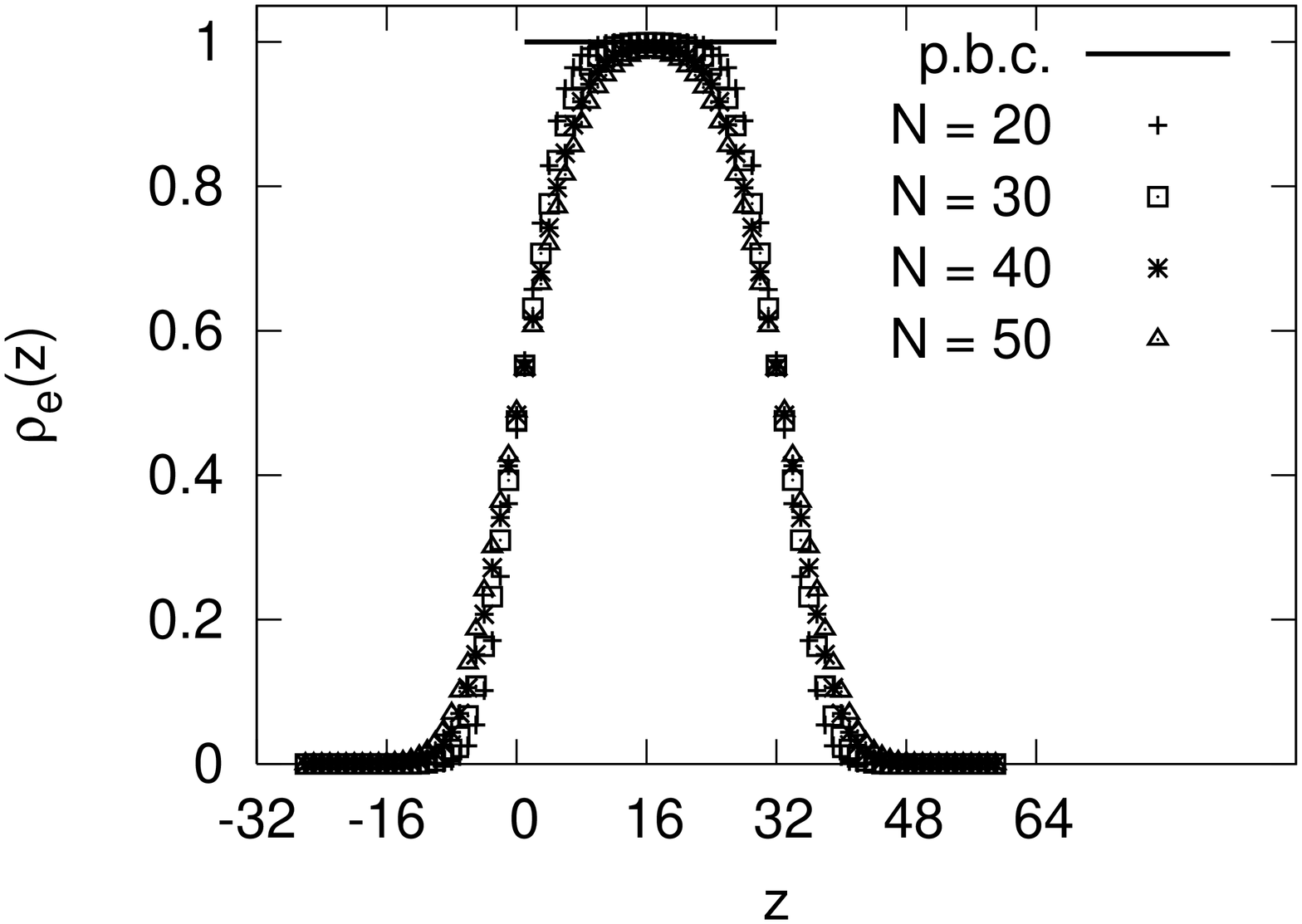}
\caption{Density distributions of all the monomers, $\rho(z)$, (a,c),
and of the  chain ends, $\rho_e(z)$ (b,d), plotted vs. $z$ for
$L_b=32$, $\sigma=1$, good solvent conditions (a,b) and $\Theta$-solvent
conditions (c,d). The distributions are normalized by choosing
$\sum_z \rho(z) =n_c$ and $\sum_z \rho_e(z)=n_c$, where
$n_c$ is the number of side chains $(n_c=\sigma L_b)$. Four chain lengths
are shown, as indicated. Note that in the p.b.c. case we trivially have
$\rho(z)=\rho_e(z)=1$, $1 \leq z \leq L_b$, for the chosen normalization.}
\label{fig5}
\end{center}
\end{figure*}

\section{STRUCTURAL PROPERTIES OF BOTTLE BRUSHES: THE EFFECT OF CHAIN
  ENDS} 
In this section we will look at the difference in structure at the
free ends of the backbone, where we can expect to find star-like
conformations for the side chains, and the central part of the
backbone which will be brush-like. Comparing
conformations for bottle-brushes with free ends to those where
pbcs are employed along the rigid backbone, we
can also find out to what extend the free ends influence the average
structure of the brush and its side chains. We will perform this
comparison for good solvent as well as for theta-solvent conditions. 

Fig.~\ref{fig1} presents our data for the perpendicular part of
the mean square gyration radius, $\langle R_{g, \bot}^2\rangle$,
where $R_{g,\bot}^2 \equiv R_{g,x}^2+R_{g,y}^2$, and the $x$ and
$y$-components refer to ``measurements'' taken in the laboratory
system with fixed orientations of the coordinate axes along the
axes of the simple cubic lattice. One sees that $\langle
R_{g,\bot}^2\rangle$ for the grafting density $\sigma =1$ is
always larger than for $\sigma =1/2$, while the dependence on
backbone length is almost invisible. In the good solvent case,
data for the decade $5 \leq N \leq 50$ are compatible with a power
law increase, but the exponent is far too small in comparison with
the prediction of the scaling theory \{for large enough $N$ 
and high grafting density one
expects~\cite{31} $\langle R_{g, \bot}^2\rangle /N^{2\nu}\propto N
^{2\nu(1-\nu)/(1+\nu)} \approx N^{0.305}$ while the effective
exponents that one can read off from Fig.~\ref{fig1}a are only
about half of this value\}. Interestingly, also in the 
$\Theta$-solvent case one observes an increase of $\langle
R_{g,\bot}^2\rangle /N^{2\nu}=\langle R_{g,\bot}^2\rangle/N$ with
increasing side chain length $N$, but there clearly occurs
curvature on the log-log plot, and thus already the data indicate
that the asymptotic region where power laws and scaling concepts
apply is not reached. Analogous data have also been taken for the
model with pbcs, but the data are almost
indistinguishable from the free end case, and hence not shown
here.

Fig.~\ref{fig2} now turns to the linear dimensions of side chains,
using a coordinate system where the $x$-direction is defined 
from the direction of the vector through the backbone and
the C.M. of each side chain, and perpendicular to the backbone direction
in each configuration (see Sec.~II), 
and also different grafting sites are distinguished, for a
relatively short backbone length, $L_b=32$. As expected
(Fig.~\ref{fig2}a), the stretching of chains grafted near the free
ends ($s_1$) in radial direction is weakest, because they acquire
a noticeable component in the $z$-direction (Fig.~\ref{fig2}c).
These effects rather quickly get weaker when the grafting site is
farther away from the chain ends, and even for a short backbone
$(L_b=32)$ the chains near the center almost behave like chains in
the bulk of a very long chain (which is modeled by eliminating
end effects through the choice of pbcs).
We do not have such an obvious interpretation for the weak (but
for the backbone ends clearly non-monotonic) variation of $\langle
R_{g,y}^2\rangle/N^{2\nu}$, however.

It is interesting to contrast these results to the $\Theta$-solvent
(Fig.~\ref{fig3}). In this case the inhomogeneity caused by the
presence of free ends of the backbone is much weaker, the
differences with respect to the p.b.c. case are much less
significant. However, a rather strong effect of the inhomogeneity
in the $z$-direction is seen when one considers the corresponding
components of the mean square backbone to end distance of the side
chains, and this effect is present both in the excluded volume
case and in the $\Theta$-solvent case (Fig.~\ref{fig4}).

\begin{figure}
\begin{center}
(a)\includegraphics[scale=0.29,angle=270]{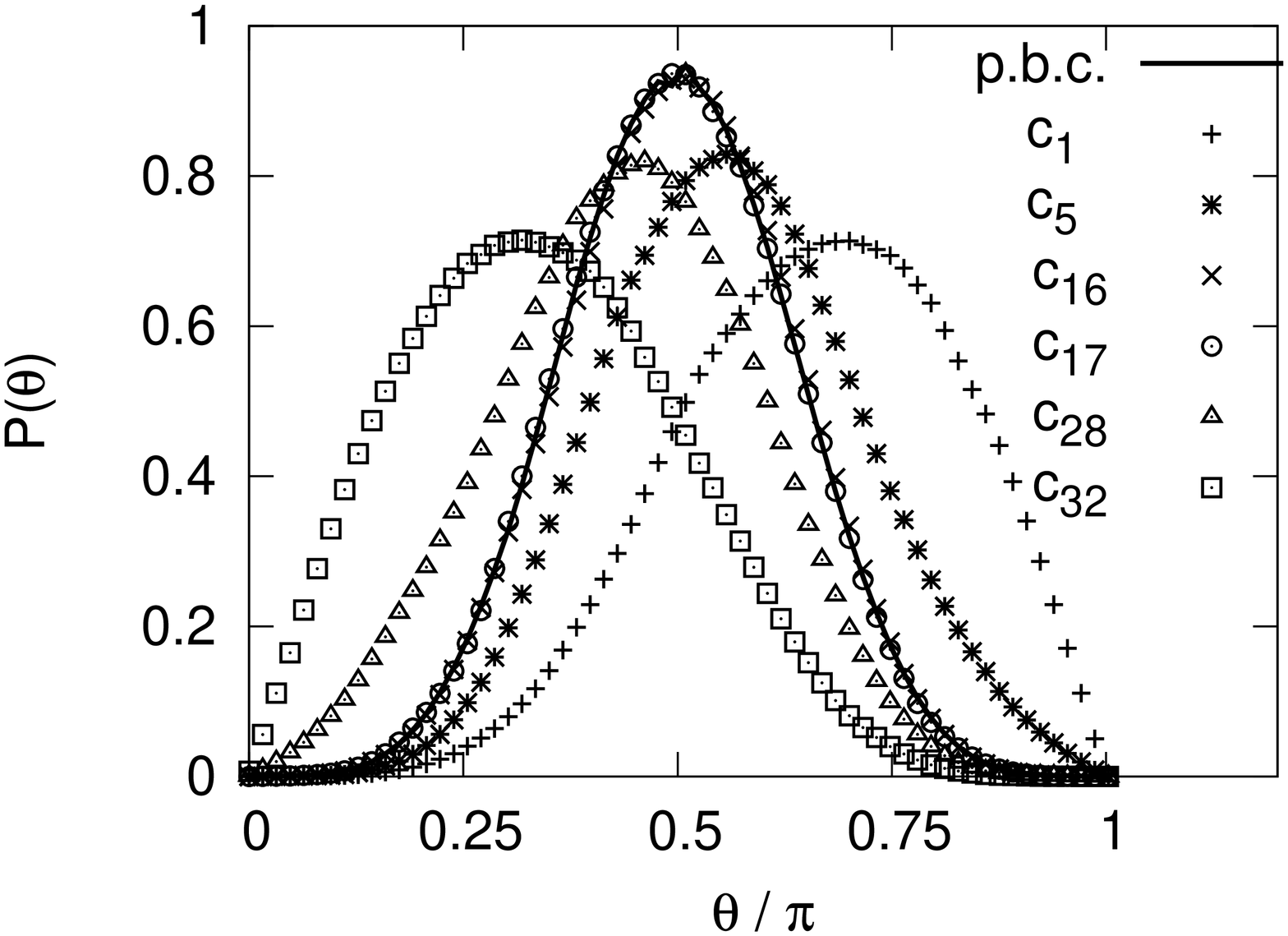}\hspace{0.4cm}
(b)\includegraphics[scale=0.29,angle=270]{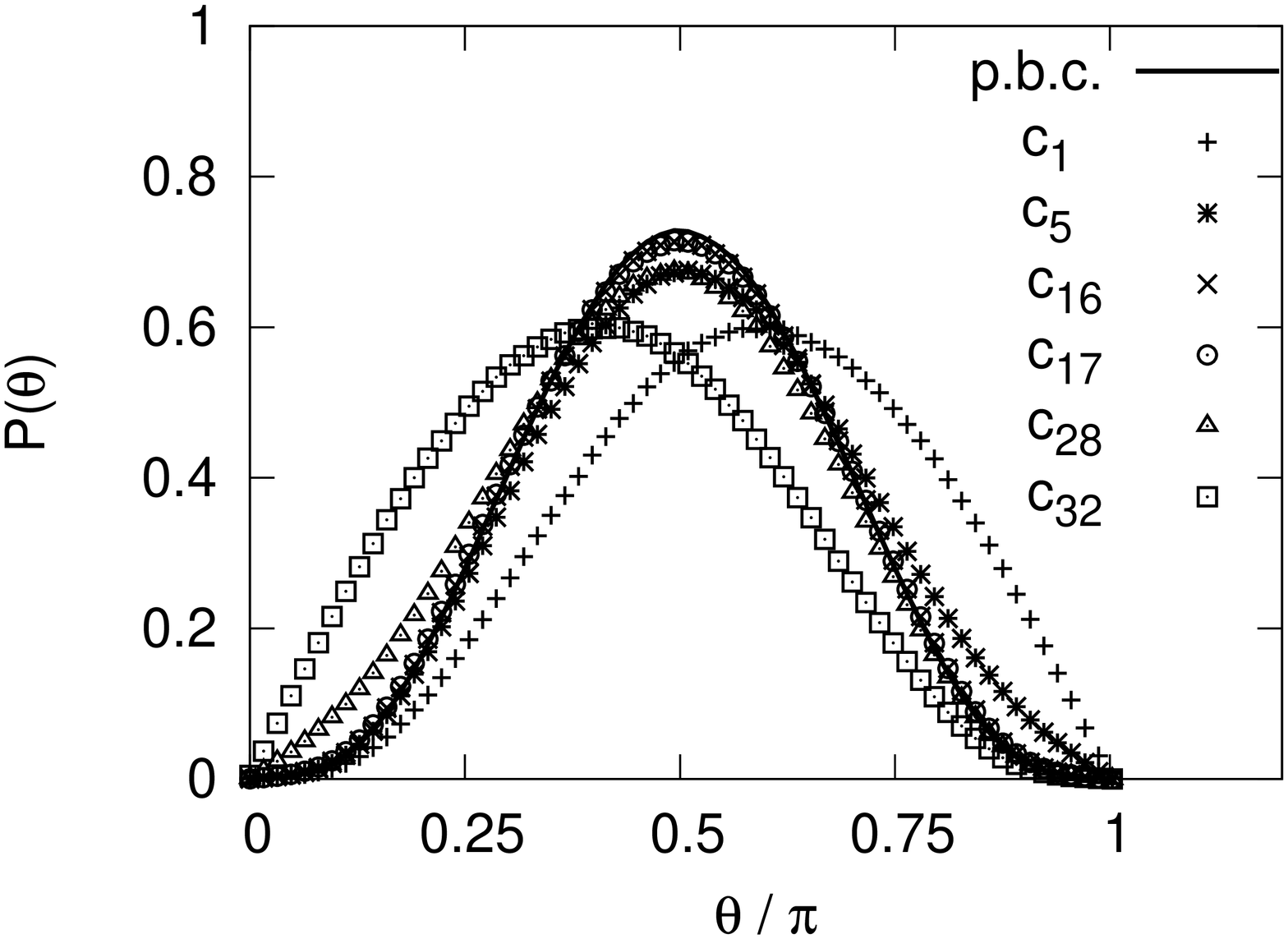}
\caption{Distribution $P(\theta)$ of
the angle $\theta$ between the vectors towards the center of mass
of each side chain and the direction of the backbone, for $L_b=32,
\sigma =1, N=50$, the good solvent case (a) and the $\Theta$-solvent
case (b). The different symbols indicate different positions along
the backbone, as indicated. The corresponding distribution for the
p.b.c. case agrees with the $c_{16}/c_{17}$ curves.}
\label{fig6}
\end{center}
\end{figure}

The next question we ask is the following: how likely is it that
monomers (or chain ends) are not in the region $1\leq z \leq L_b$
where the grafting sites are? Fig.~\ref{fig5} shows also from this
criterion that in the good solvent case the bottle-brush is more
extended in the $z$-direction than in the $\Theta$-solvent case. Even
for short backbones ($L_b=32$) for $\Theta$-solvents bulk behavior
is reached, while for good solvents there is still some systematic
depression in the center $(z=L_b/2=16)$. We note, however, that
for larger $L_b$ such as $L_b=64$ (to save space
these data are not shown) bulk behavior is reached for a
significant range of $z$ in the center of the bottle-brush.

Another quantity that shows that side chains near the backbone
ends tend to orient much more along the $z$-axis in the good solvent
case rather than in the $\Theta$-solvent case is the distribution
$P(\theta)$ of the angle between the vectors towards the center of
mass of each side chain and the $z$-direction (Fig.~\ref{fig6}). One
should note that angles $\theta$ near $\theta = \pi/2$ characteristic
for chains stretched away from the backbone in perpendicular
direction, dominate only in the center of the backbone, while
angles near $\theta =\pi/4$ and $3\pi/4$ make a substantial
contribution near the backbone ends. For the considered side chain
length, this effect dies out after a few monomeric distances away
from the backbone ends, however. For $\Theta$-conditions
(Fig.~\ref{fig6}b) this behavior is only found close to the chain
end. Chains grafted already five monomers away from the backbone end
show no tilting like for the good solvent case.
The average angle remains at $\pi/2$, but
the distribution gets broader and asymmetric with a heavy
tail towards the adjacent chain end.

\begin{figure}
\begin{center}
(a)\includegraphics[scale=0.29,angle=270]{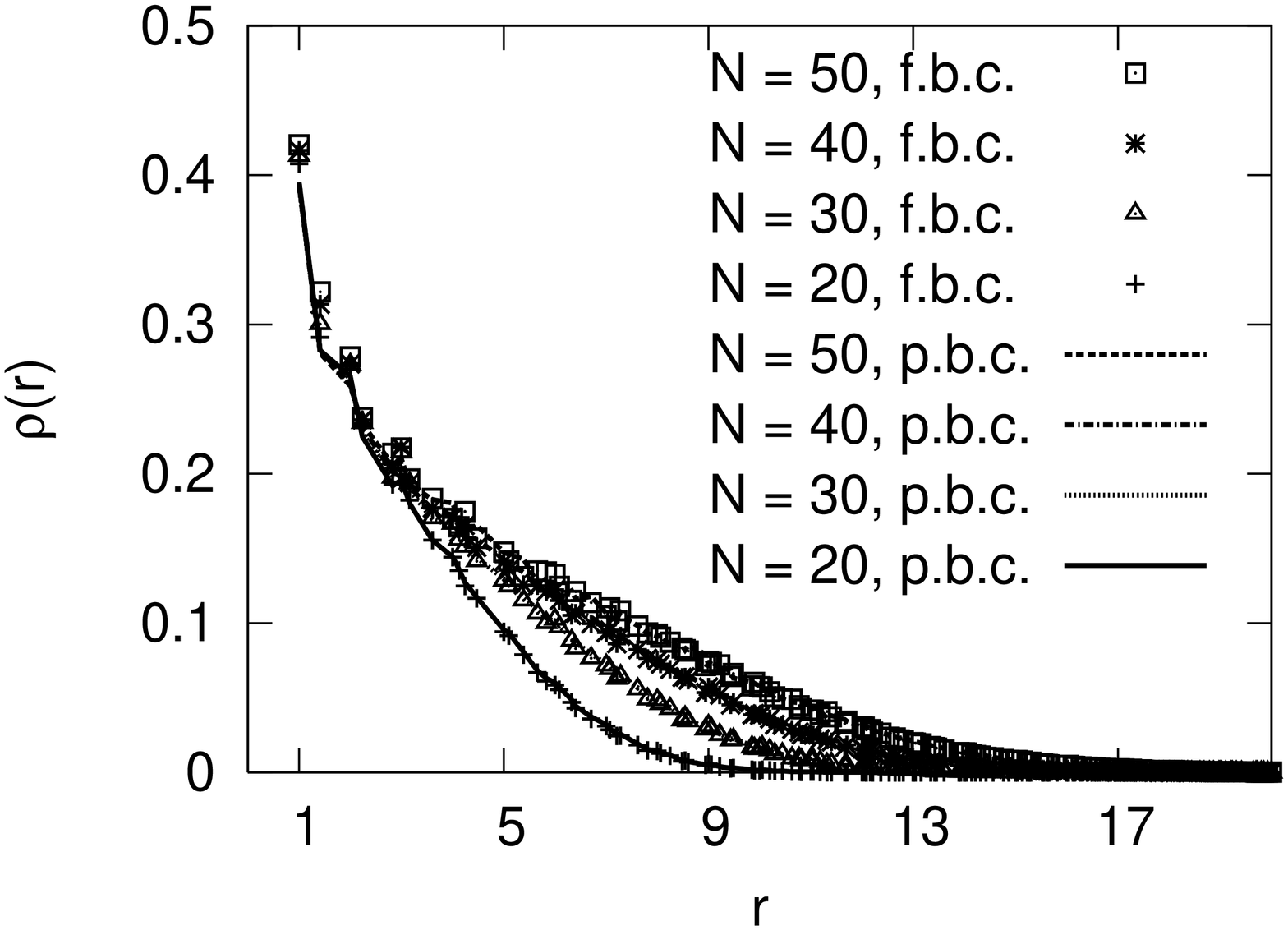}\hspace{0.4cm}
(b)\includegraphics[scale=0.29,angle=270]{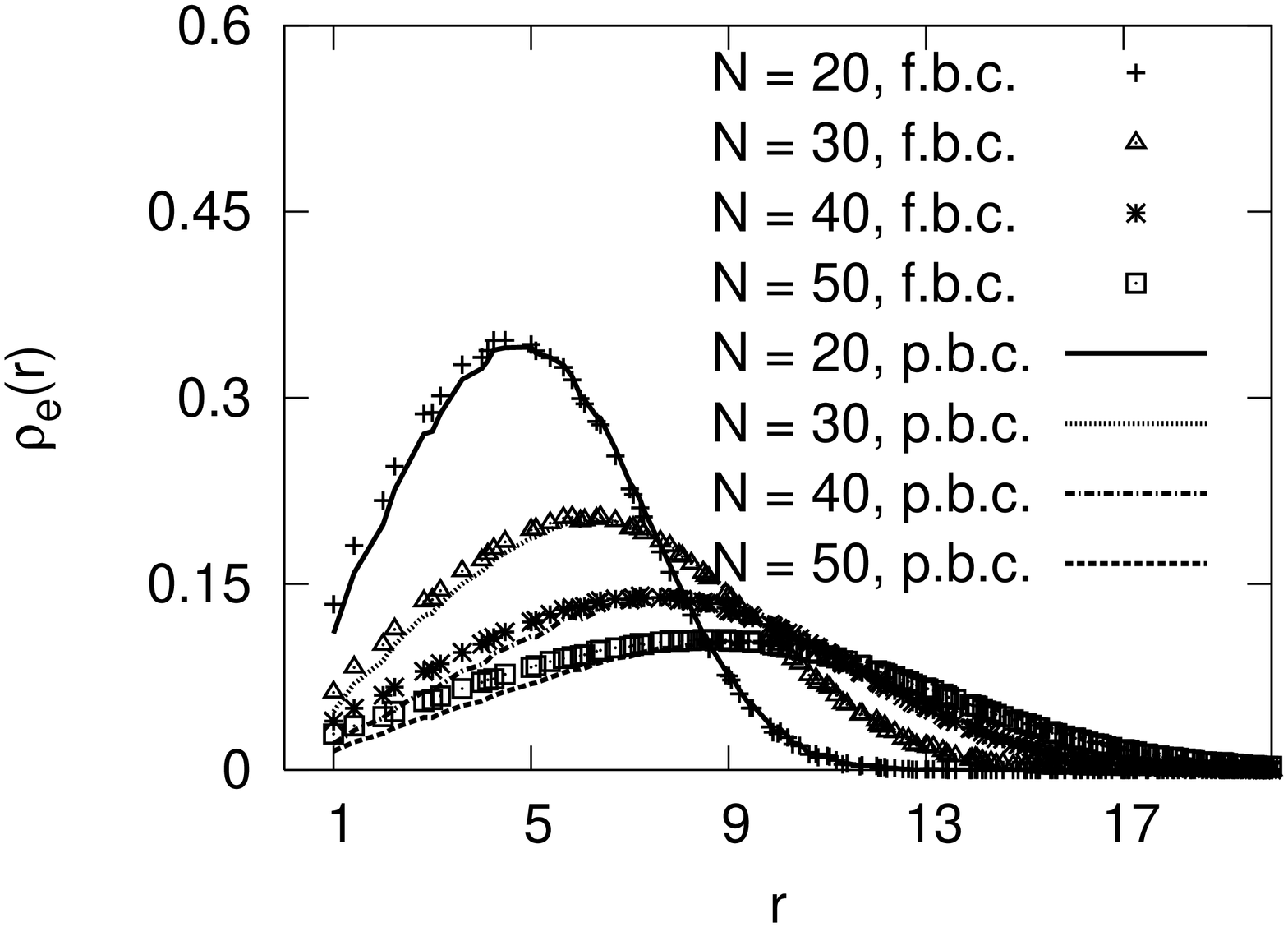}
\caption{Radial distribution function
$\rho(r)$ of all monomers (a) and radial distribution function
$\rho_e(r)$ of chain ends (b), plotted versus $r$ for $L_b=64,
\sigma =1$, good solvent conditions, and four values of the side
chain length N, as indicated. Symbols show our results for free
ends, while curves show corresponding data for the case of pbcs.}
\label{fig7}
\end{center}
\end{figure}

\begin{figure}
\begin{center}
(a)\includegraphics[scale=0.29,angle=270]{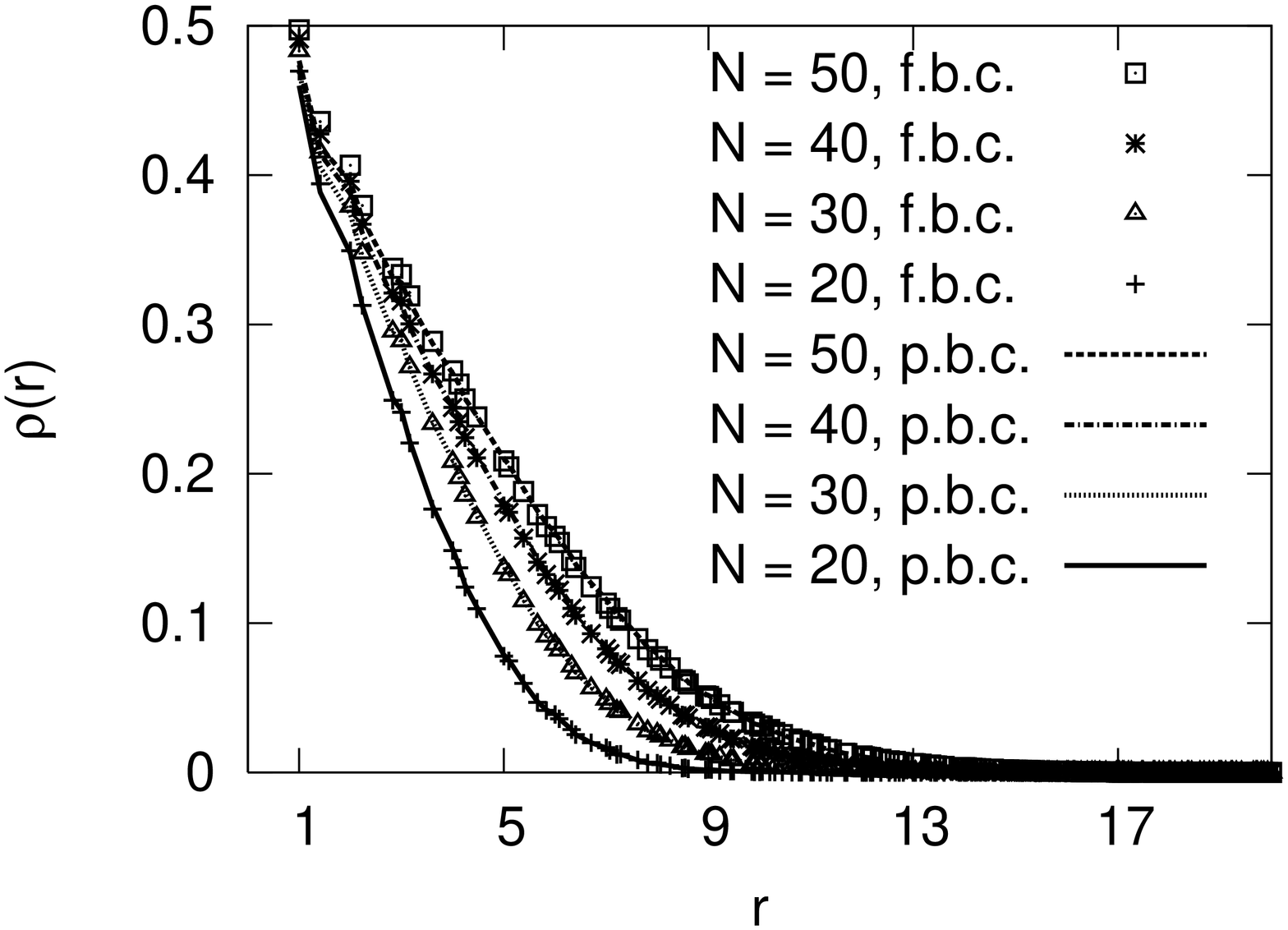}\hspace{0.4cm}
(b)\includegraphics[scale=0.29,angle=270]{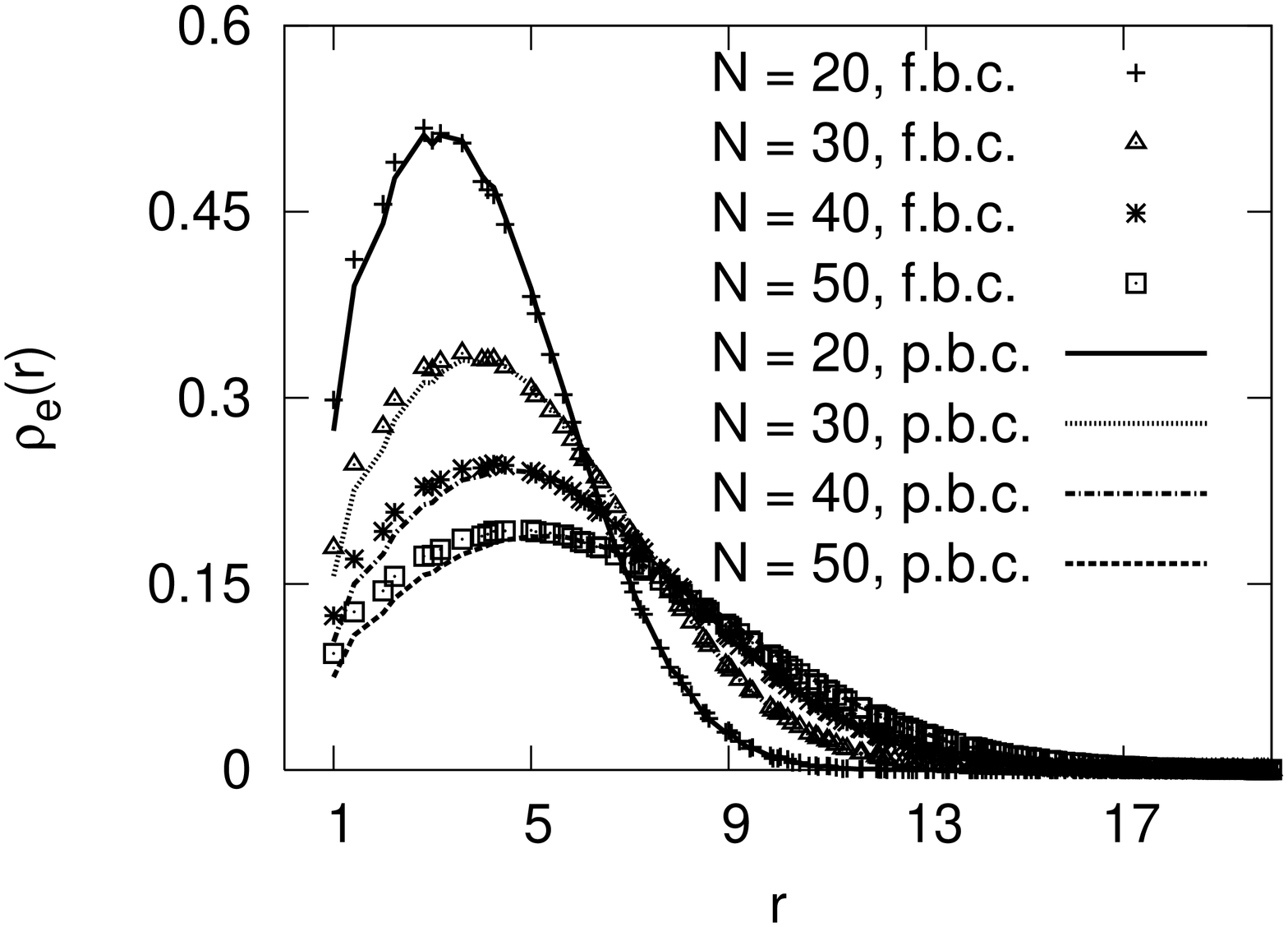}
\caption{Same as Fig,.~\ref{fig7}, but
for $\Theta$-solvent conditions.}
\label{fig8}
\end{center}
\end{figure}

The data shown in Figs.~\ref{fig2}-\ref{fig6} are readily
accessible in simulation, but not easy to access experimentally.
They help, nevertheless, to develop a complete picture of the
structure of bottle-brush polymers and clarify the side chain
conformations. Quantities, that experimentalists try to extract
from their studies are accessible to the simulations as well, of
course. Such quantities are the radial distribution $\rho(r)$ of
the monomers and $\rho_e(r)$ of the chain ends 
(shown in Figs.~\ref{fig7}, \ref{fig8} for $L_b=64$). 
Note that due to the discreteness of the lattice, the
number of monomers, $N(r)$ and $N_e(r)$, in the interval 
$[r, r + dr]$ are normalized i.e. $\rho(r)= N(r)/N_r$ and
$\rho_c(r)=N_e(r)/N_r$ where $N_r$ is the number of lattice
points with a distance to the backbone lying in the interval $[r,r+dr]$. 
For comparing data of different chain
lengths, normalization conditions $\sum \limits _r N(r)=N$ and
$\sum \limits _r N_e(r)=n_c$ have been imposed. Similar data have
also been generated for $L_b=32$, but the differences to those
shown are only small, and therefore need not be discussed here.
Figs.~\ref{fig7} and \ref{fig8} reveal that neither $\rho(r)$ nor
$\rho_e (r)$ are sensitive to the effects of the free ends: for
the chosen $L_b$, much longer side chains would be required in
order that effects due to the crossover from bottle-brush to
star polymer behavior come into play. While in the good solvent case the
chain ends are typically farther away from the backbone than in
the $\Theta$-solvent case, the qualitative behavior of $\rho(r)$ and
$\rho_e(r)$ does not depend on solvent quality much. Furthermore,
it is gratifying to note that these data are qualitatively rather
similar to the corresponding Molecular Dynamics results of Murat
and Grest~\cite{11} for a bead-spring off lattice model of
flexible side chains tethered to a straight line. This similarity
reinforces our view that on a coarse grained level, the present
lattice model should yield useful results.

\begin{figure}
\begin{center}
(a)\includegraphics[scale=0.29,angle=270]{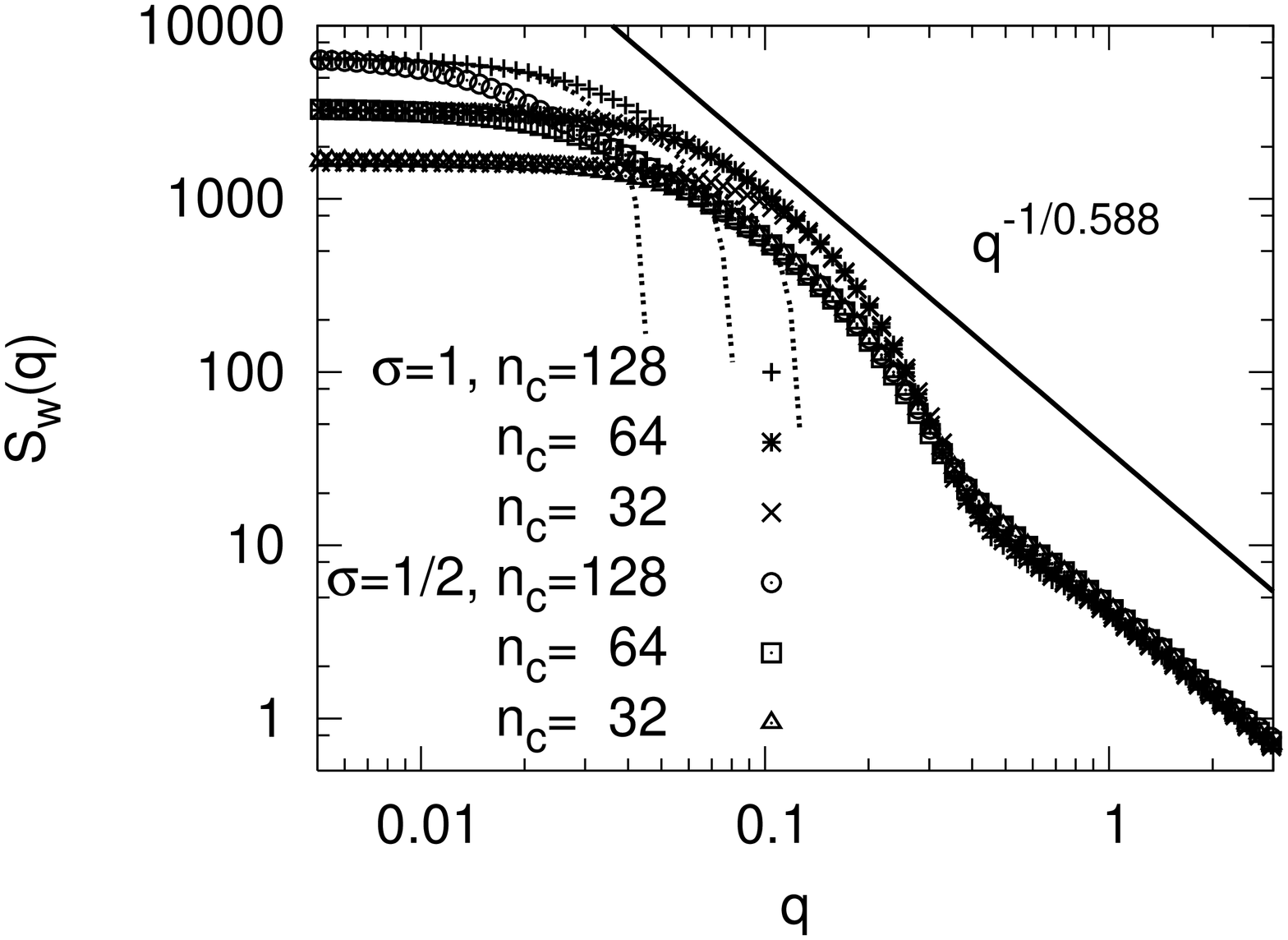}
\hspace{0.4cm}
(b)\includegraphics[scale=0.29,angle=270]{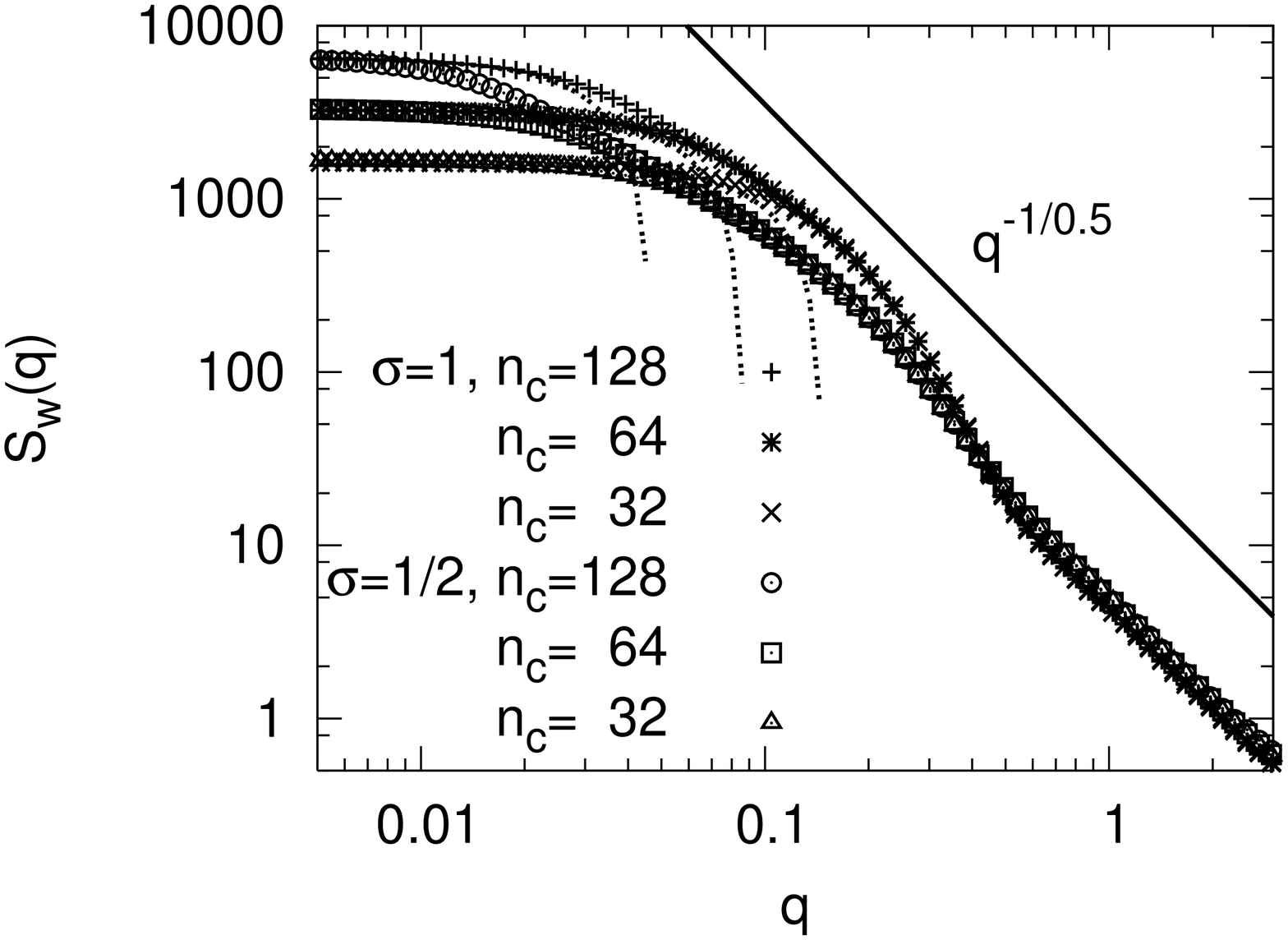}
\caption{Log-log plot of the scattering
function of the whole bottle-brush polymer, $S_{\rm w}(q)$, in a good
solvent (a) and a $\Theta$-solvent (b) versus $q$. All data are for
the case of free ends, $N=50$, while data for two choices of
$\sigma$ and three choices of $n_c$ each are included, as
indicated. Straight lines show the theoretical power laws for
flexible chains and intermediate $q$-values, $S_{\rm w}(q)\propto
q^{-1/\nu}$ and (a) $\nu = 0.588 $ or (b) $\nu=0.5$ (b), respectively.
Dotted curves are given by Eq.~(\ref{eq2})
for $\sigma=1$.}
\label{fig9}
\end{center}
\end{figure}

\section{SCATTERING FUNCTIONS FOR BOTTLE-BRUSH POLYMERS AND THEIR
THEORETICAL MODELING}

Let us now turn to a discussion of experimentally observable
information on the structure of a bottle brush polymer. In experiments
one has to infer the structure from scattering data \cite{2,3,4}
employing suitable model assumptions on the structure. In the
simulation we obtain both, the scattering data and the underlying
structural properties described in the last section independently, and
therefore are able to test theoretical models suggested to link the two. 
Fig.~\ref{fig9} presents our data for the total scattering
function $S_{\rm w}(q)$ for the bottle-brush polymers, both for good
solvent and $\Theta$-solvent conditions. Here $S_{\rm w}(q)$ is defined as

\begin{equation}\label{eq1}
S_{\rm w}(q) = \frac {1}{\mathcal{N}_{\textrm{tot}}} \sum \limits
_{i=1}^{\mathcal{N}_{\textrm{tot}}}
\sum\limits_{j=1}^{\mathcal{N}_{\textrm{tot}}} \langle
c(\vec{r}_i)c(\vec{r}_j)\rangle \frac {\sin (q|\vec{r}_i-
\vec{r}_j|)}{q|\vec{r}_i-\vec{r}_j|},
\end{equation}
where $c(\vec{r}_i)$ is an occupation variable, $c(\vec{r}_i)=1$ if
the site $\vec{r}_i$ is occupied by a bead, and zero otherwise. Note
that an angular average over the direction of the scattering
vector $\vec{q}$ has been performed, and the sums run over all
monomers (all side chains and the backbone).

Surprisingly, our data are qualitatively very
similar to the corresponding experimental data (see e.~g.~Fig.~\ref{fig4} 
of~\cite{2}), although the latter refer to a
polymer with a flexible backbone, unlike our simulations.
As always, the limit $q \to 0$ of $S_{\rm w}(q)$ reflects the total
number $\mathcal{N}_{\textrm{tot}}$ of scattering monomers,
and the leading deviation from it is described by the total
gyration radius,

\begin{equation}\label{eq2}
S_{\rm w}(q) \approx \mathcal{N}_{\textrm{tot}}[1-q^2\langle
R_g^2\rangle /3]\;.
\end{equation}
This behavior is shown by the fine-dotted lines in
Fig.~\ref{fig9} for the case of grafting density $\sigma=1$.
The $q$ range over which this approximation agrees with the scattering
data increases with increasing ratio of side chain
length to backbone length, $N/L_b$.
Of course, more
interesting is the behavior at larger $q$, where Eq.~(\ref{eq2}) is
no longer valid. The region where $S_{\rm w}(q)$ is strongly curved and
decreases rapidly ($0.1 \leq q \leq 0.3$, in our case) has
contributions from the conformation along the backbone (rigid rod in
our case which should show up as a behavior $S{\rm w}(q) \simeq
q^{-1}$ for longer backbones) and from the scattering from the cross section
through the cylindrical bottle-brush, and needs to be related to
data such as shown in Figs.~\ref{fig7}, \ref{fig8}. The $q$
range near $q=1$ reflects the self-avoiding walk structure
$q^{-1/\nu}$ before it is affected by the
local packing of monomers on the lattice at still larger $q$, and in real systems
reflects local properties such as the persistence length of the
flexible side chains, possible scattering from side groups, etc.
This non-universal regime hence is less interesting. From this
discussion of the total structure factor we can already conclude that
it is the $q$-range $0.04 \le q \le 0.5$ which for our model contains
the important information about the structure of the brush.

\begin{figure*}
\begin{center}
(a)\includegraphics[scale=0.29,angle=270]{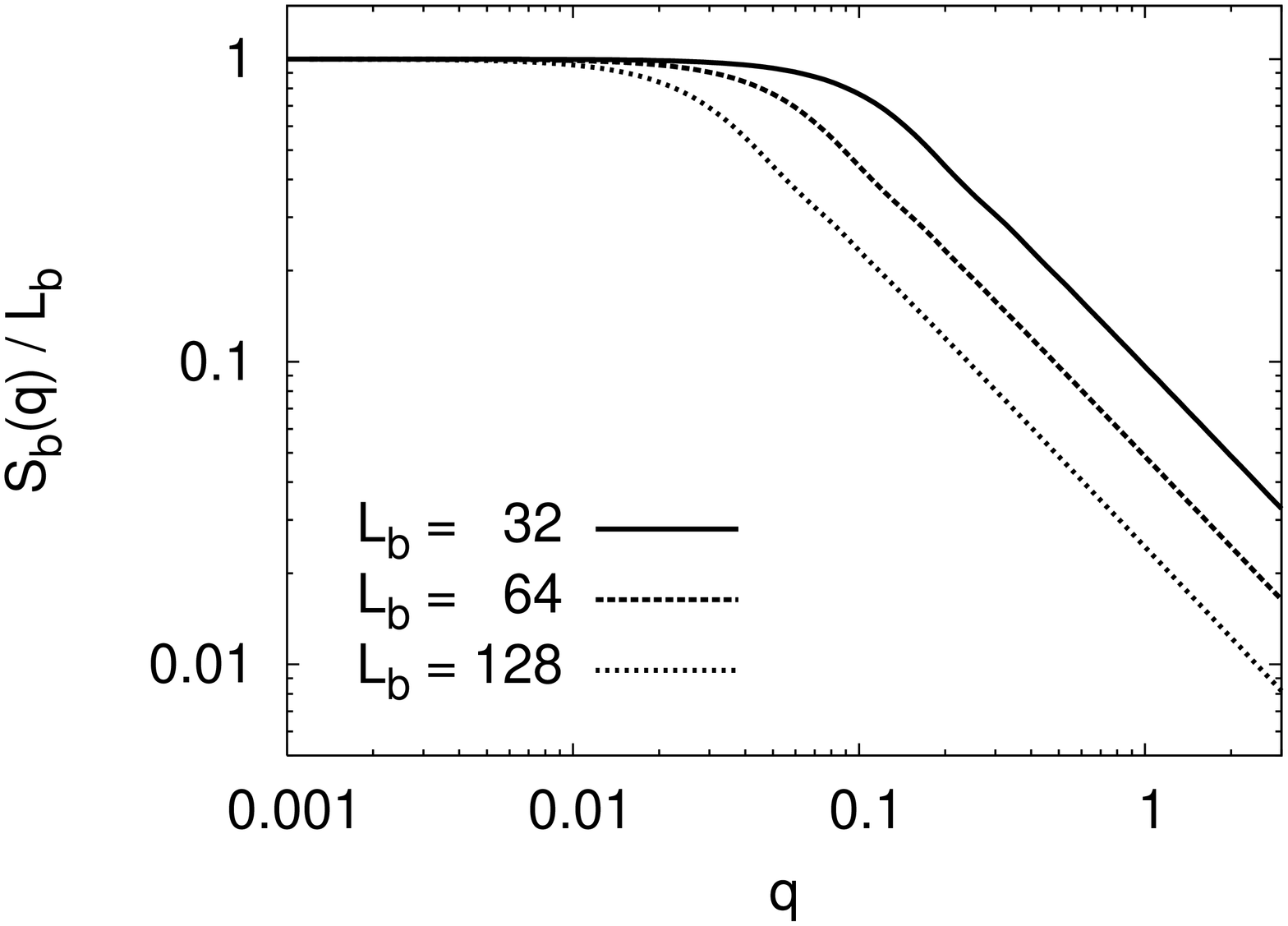}\\
(b)\includegraphics[scale=0.29,angle=270]{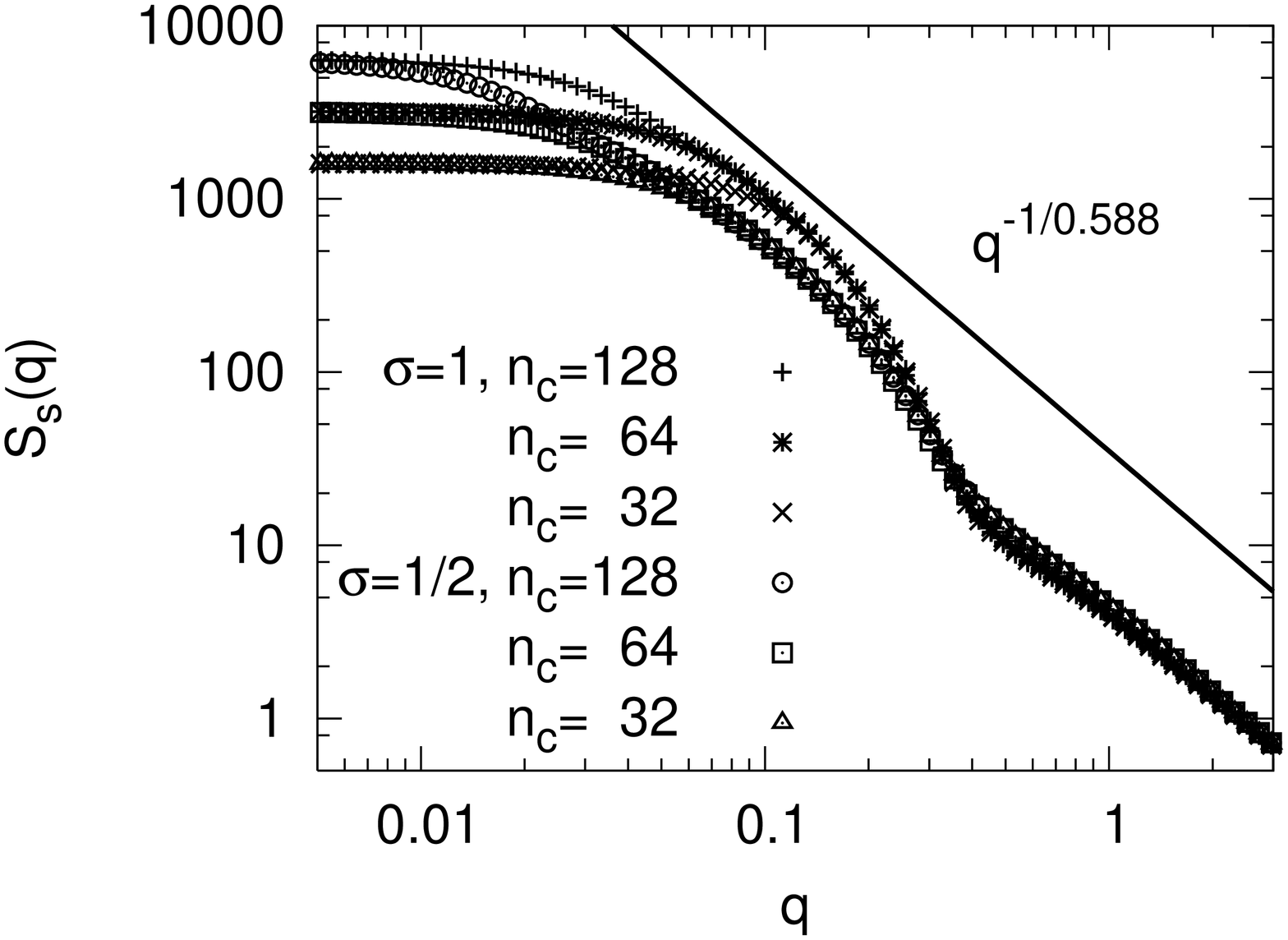}\\
(c)\includegraphics[scale=0.29,angle=270]{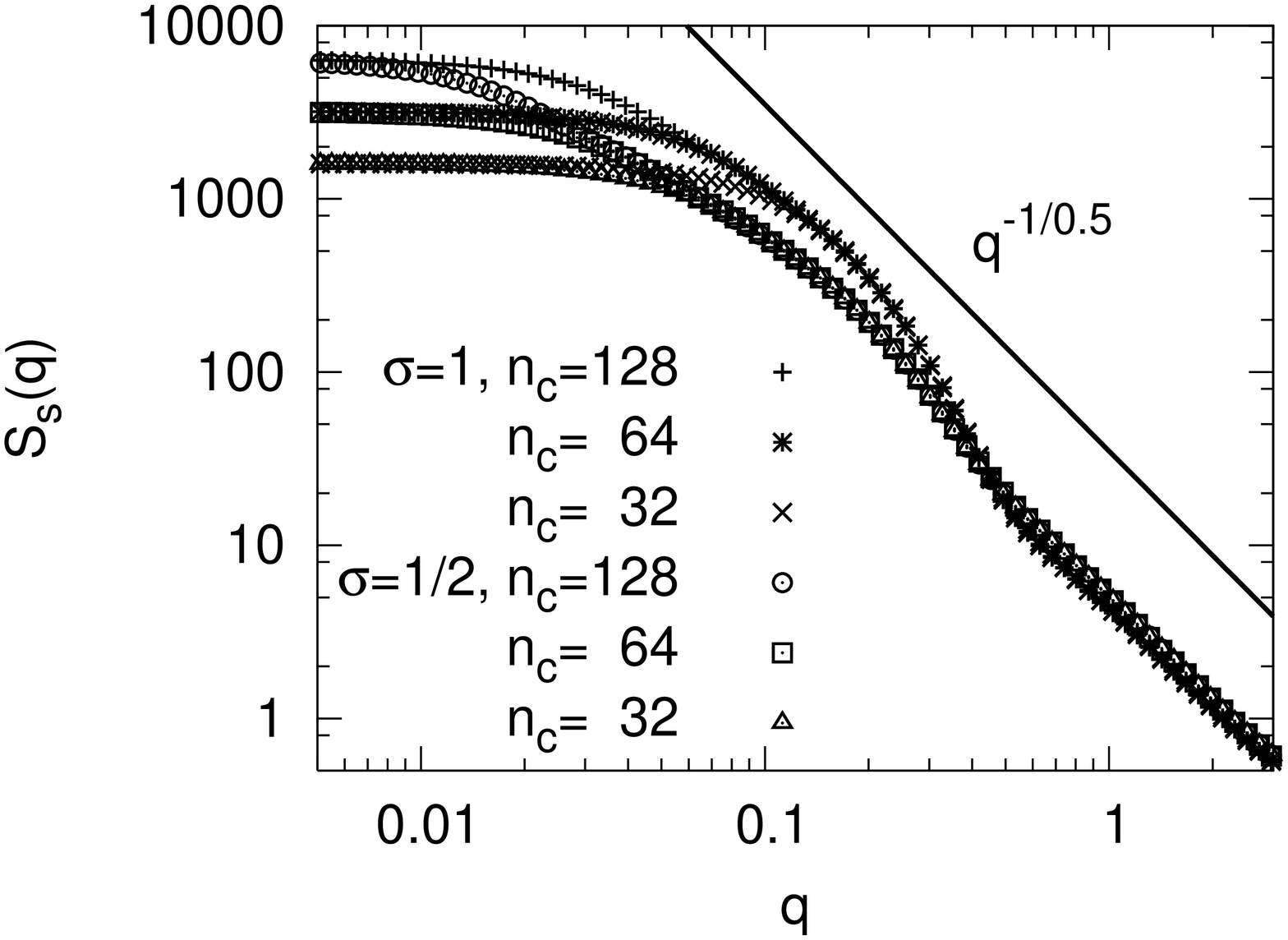}
\caption{(a) Log-log plot of the normalized scattering function of the backbone,
$S_{\rm b}(q)/L_b$ versus $q$, using the formula of Pedersen and
Schurtenberger~\cite{39}, see Eq.~(\ref{eq5}) for the case of thin rods.
(b) Log-log plot of
the scattering from all monomers in the side chains of the bottle-brush
with $N=50$. The data are for good solvent conditions, and for two
choices of $\sigma$ and three choices of $n_c$ each, as indicated.
Straight lines have the same meaning as in Fig.~\ref{fig9}.
(c) Same as (b), but for $\Theta$-solvent conditions.
Note that $S_s(q)$ is normalized such that $S_s(q=0)=Nn_c$.}
\label{fig10}
\end{center}
\end{figure*}

One advantage of our simulations is that we can obtain scattering
contributions from different parts separately. E.~g., we can
isolate the scattering from the backbone (Fig.~\ref{fig10}a) and from
the scattering of the side chains (Fig.~\ref{fig10}b,c). In
our case, where the backbone is a rigid rod where just the
subsequent sites $i=1,2,\ldots, L_b$ are taken by monomers,
Eq.~(\ref{eq1}) becomes

\begin{equation}\label{eq3}
S_{\rm b}(q) = \frac {1}{L_b} \sum \limits
_{i=1}^{L_b} \sum \limits ^{L_b}_{j=1} \frac{\sin (q|j-i|)}{q|j-i|} \;.
\end{equation}
Noting that the distance $|j-i|=0$ occurs $L_b$ times, while the
distance $|j-i|=1$ occurs $2(L_b-1)$ times, the distance $|j-i|=2$
occurs $2(L_b-2)$ times, etc., we conclude that

\begin{equation}\label{eq4}
S_{\rm b}(q) =-1+ \frac{2}{L_b} \sum \limits _{k=0}^{L_b-1}
(L_b-k) \frac{\sin(qk)}{qk} \;.
\end{equation}
The factor 2 accounts for the fact that both positive
and negative differences $k=j-i$ occur, and the extra $-1$
corrects for over counting in the term $k=0$.

In the limit where $L_b\rightarrow \infty$ and $qL_b$ is of order
unity, the sum in Eq.~(\ref{eq4}) can be transformed into an
integral, to find

\begin{equation}\label{eq5}
S_{\rm b}(q) \approx \frac{2}{q} \int _0^{qL_b} \frac{\sin t}{t} dt
-4 \frac{\sin ^2(qL_b/2)}{q^2L_b}
\end{equation}
Eq.~(\ref{eq5}) is nothing but the well-known scattering function of
an infinitely thin rod of length $L_b$ with a continuous mass
distribution along the rod~\cite{39,40}.

\begin{figure}
\begin{center}
(a)\includegraphics[scale=0.29,angle=270]{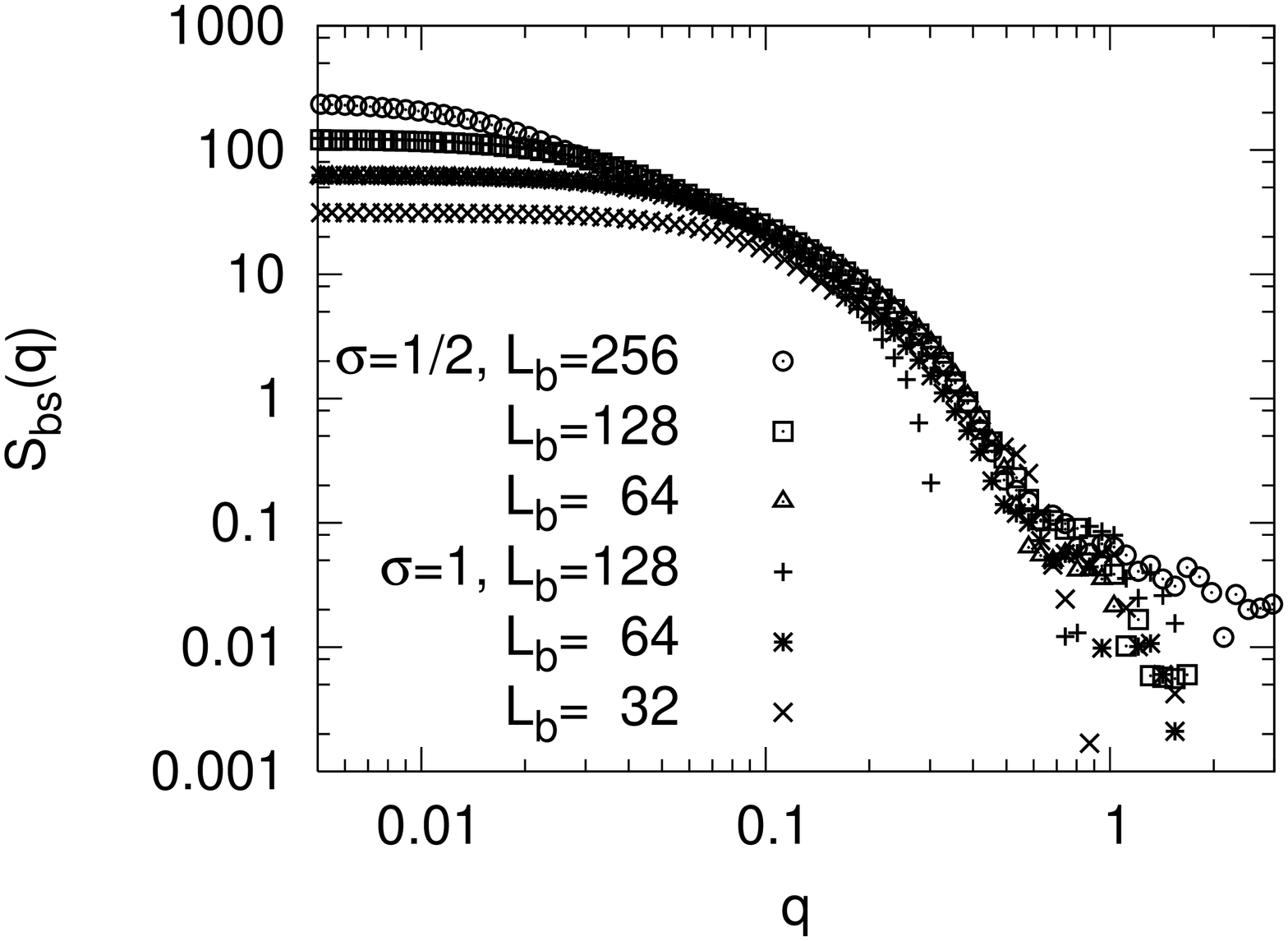}
\hspace{0.4cm}
(b)\includegraphics[scale=0.29,angle=270]{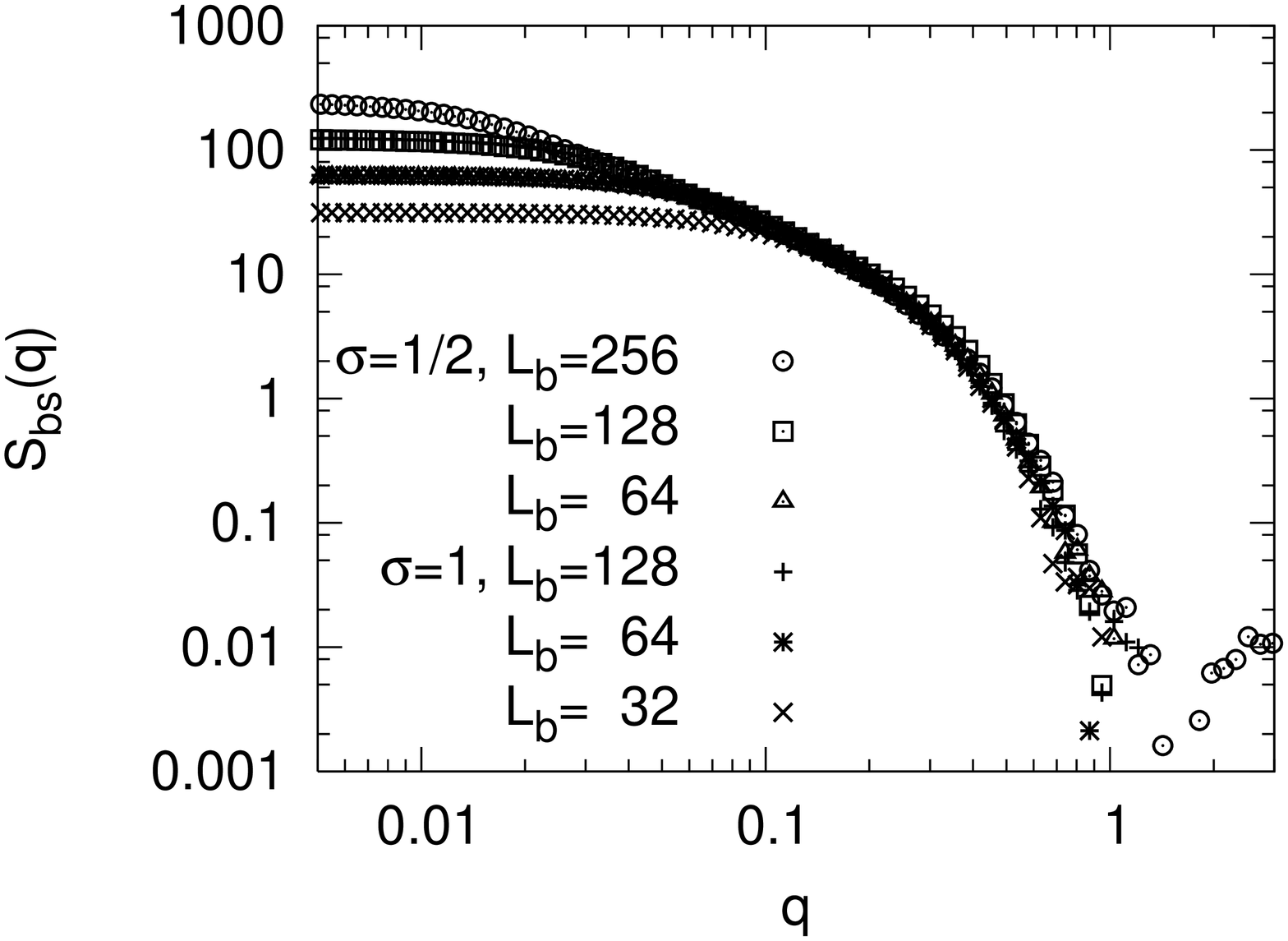}
\caption{
$S_{\rm bs}(q)=[N_{\rm tot}S_{\rm w}(q)-L_bS_b(q)-Nn_cS_s(q)]/(2N_{\rm tot})$,
case (a) is for good solvent conditions, case (b) for $\Theta$-solvent
conditions.}
\label{fig11}
\end{center}
\end{figure}

According to Eq.~(\ref{eq1}), the scattering function of 
all side chains is given by

\begin{equation}\label{eq5a}
S_{\rm s}(q) = \frac {1}{Nn_c} \sum \limits
_{i=1}^{Nn_c}
\sum\limits_{j=1}^{Nn_c} \langle
c(\vec{r}_i)c(\vec{r}_j)\rangle \frac {\sin (q|\vec{r}_i-
\vec{r}_j|)}{q|\vec{r}_i-\vec{r}_j|} \;.
\end{equation}
When we add $S_{\rm b}(q)$ and $S_{\rm s}(q)$ with their relative weights, 
see Figs.~\ref{fig10}a,b, we do not recover $S_{\rm w}(q)$
strictly, due to interference effects in the scattering from
monomers in the side chain and in the backbone. Such interference
effects normally are neglected~\cite{2,3,4,4a}. Taking the difference
$S_{bs}=[N_{\rm tot}S_{\rm w}(q)-L_bS_b(q)-Nn_cS_s(q)]/(2N_{\rm tot})$
, we can test for the
importance of such interference effects as shown in Fig.~\ref{fig11}. 
Indeed, we do find that such
interference effects are present although only at a percent level.

We now turn to a discussion of the scattering from the side
chains, which clearly dominates the scattering intensity in all
cases of practical interest. For scattering wavenumber $q$ in the
range $q \langle R_g^2\rangle \gg 1$ this scattering should be
dominated by the cross sectional structure of the bottle-brush. 
In the analysis of the experimental scattering data one has to assume that 
one can determine the cross-sectional contribution by a factorization

\begin{equation}\label{eq8}
S_{\rm w}(q)\equiv S_{\rm b}(q)S_{\rm xs}(q)\;,
\end{equation}
where $S_{\rm xs}(q)$ is interpreted as the cross section structure
factor. Such decoupling approximations seem to be successful
for worm-like micelles~\cite{39}. In the literature, $S_{\rm b}(q)$
is modeled by a superposition of rigid rod and worm-like chain form factors,
needed to account for backbone bending~\cite{4a}. In our case we can take
Eq.~(\ref{eq8}) simply as a {\it definition} of $S_{\rm xs}(q)$ using $S_{\rm b}(q)$
which is known exactly for our case (see Eq.~(\ref{eq4})).

The cross sectional scattering is then assumed to be obtainable from a
rotationally averaged two-dimensional Fourier transform of the radial density
distribution. 

\begin{equation}
S_{\rm xs}(q) = \frac{1}{C} \langle \left| \int d^2\vec{r} \rho(\vec{r})
{\rm e}^{i\vec{q}\cdot\vec{r}} \right|^2\rangle_{\rm T,\hat{q}}.
\end{equation}
Here $C$ is a normalization, and the indices $T$ and $\hat{q}$ indicate a
thermal average and an average over the unit circle in two dimensions. This
is further approximated by neglecting correlations in the radial density fluctuations 

\begin{equation}
\langle \rho(\vec{r})\rho(\vec{r}\thinspace')\rangle_{\rm T} =  \langle
\rho(\vec{r})\rangle_{\rm T}\langle\rho(\vec{r}\thinspace')\rangle_{\rm T} =: \rho(r) \rho(r')
\end{equation} 
to obtain

\begin{equation}
S_{\rm xs}(q) = \frac{1}{C} \left| \int d^2\vec{r} \rho(r) \langle
{\rm e}^{i\vec{q}\cdot\vec{r}}\rangle_{\rm \hat{q}} \right|^2.
\end{equation}
With the proper normalization this yields

\begin{equation}\label{sxc-final}
S_{\rm xs}(q) = \frac{\left| \int_0^\infty dr r \rho(r) J_0(qr)
  \right|^2}{\left| \int_0^\infty dr r \rho(r) \right|^2},
\end{equation} 
where $J_0(r)$ is the zeroth order Bessel function of the first kind.
With the approximations underlying Eq.(\ref{sxc-final}) the experimental cross
section structure factor can be inverted to obtain the radial density
distribution

\begin{equation}\label{eq10}
\rho_{\rm xs}(r)= \frac {1}{2 \pi} \int \limits _0^\infty [S_{\rm xs}(q)]
^{1/2} J_0(qr)qdq.
\end{equation}

In the analysis of experimental data, different plausible assumptions for
the radial density profile were used, guided by the assumed similarity to the
scattering from worm-like micelles. Rathgeber et al.~\cite{2} propose to use the
following empirical function

\begin{eqnarray}\label{eq6}
g(r) = \left\{ \begin{array}{l c r}
 1 & \textrm{for}&\; r \leq R_c \; \\
\alpha r^{-k} \{1+\exp[(r-R_s)/\sigma _s]\}^{-1} \; &
\textrm{for}&\; r >R_c
\end{array} \right.
\end{eqnarray}
Here $R_c$ is an inner radius, up to which $\rho(r) $ is a constant;
then there is a power law decay, described by an exponent $k$, up to
some outer radius $R_s$, then a fast decay to zero (over the range
$\sigma_s$) follows. The constant $\alpha$ is fixed by the condition that
$g(r=R_c)$ is continuous, so Eq.~(\ref{eq6}) involves the four
nontrivial fitting parameters $R_c$, $k$, $R_s$ and $\sigma_s$. 
Zhang et al.~\cite{4} assume a form for the cross section structure factor
in terms of the first order Bessel function $J_1(x)$,

\begin{equation}\label{eq11}
S_{\rm xs}(q)= {\rm const} [\frac {2J_1(qR_c)}{qR_c} \exp (-q^2s^2/2)]^2\;,
\end{equation}
where $R_c$ is an effective radius, and $s$ is an effective width. 
This
is equivalent to assuming a radial density profile which is a convolution of a
step function and a Gaussian

\begin{equation}
\rho_{\rm xs}(r)= \rho_0 \int d^2\vec{r}\thinspace'
\left[1-\theta(|\vec{r}\thinspace'|-R_c)\right]\exp\left[-\frac{(\vec{r}-\vec{r}\thinspace')^2}{2s^2}\right].
\end{equation}
Here we have three free parameters, $\rho_0, R_c$ and $\sigma$.
Again, $R_c$ is a measure of the range over which the density profile is assumed to be
flat in the core of the bottle-brush  

\begin{figure}
\begin{center}
(a)\includegraphics[scale=0.29,angle=270]{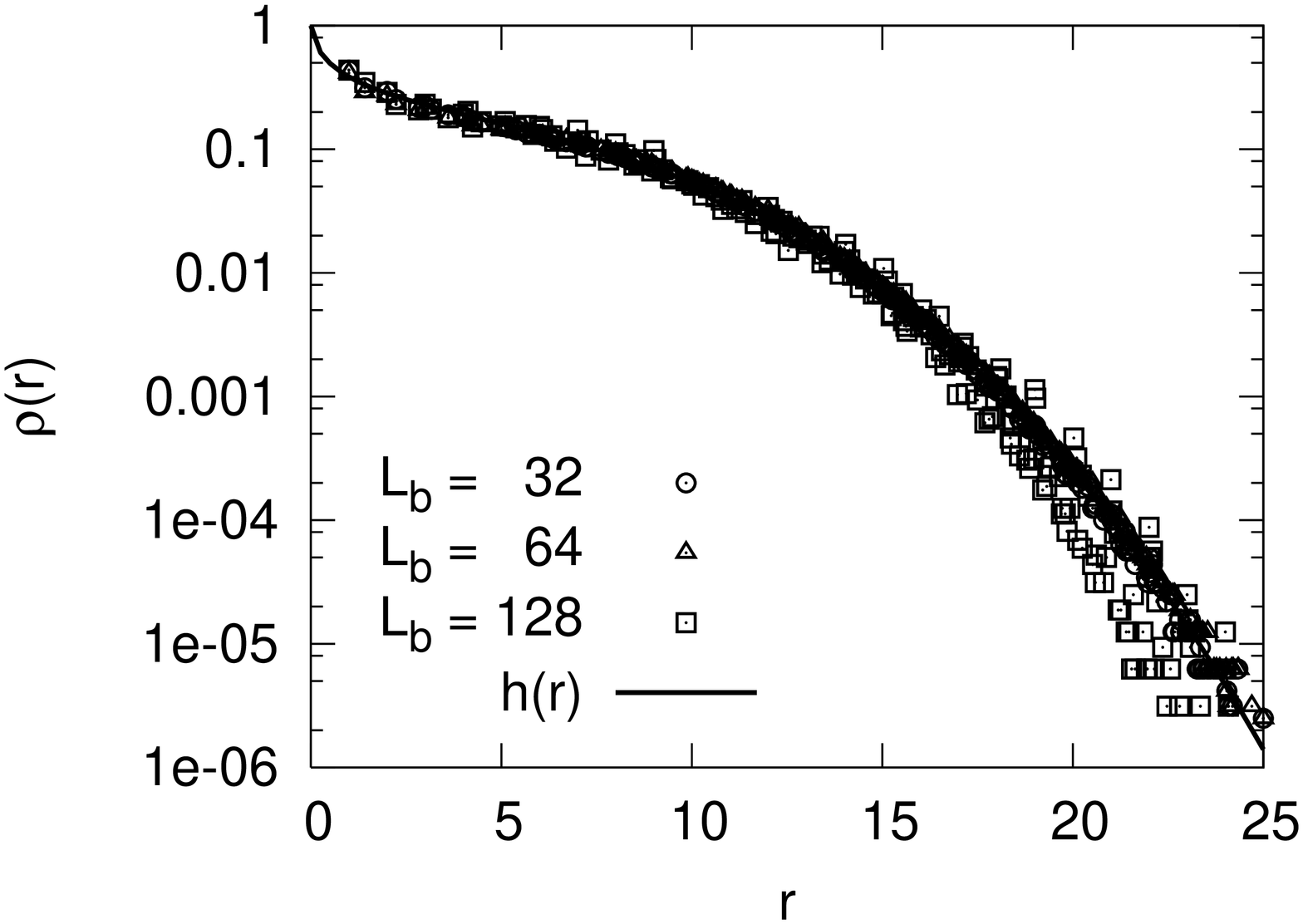}
\hspace{0.2cm}
(b)\includegraphics[scale=0.29,angle=270]{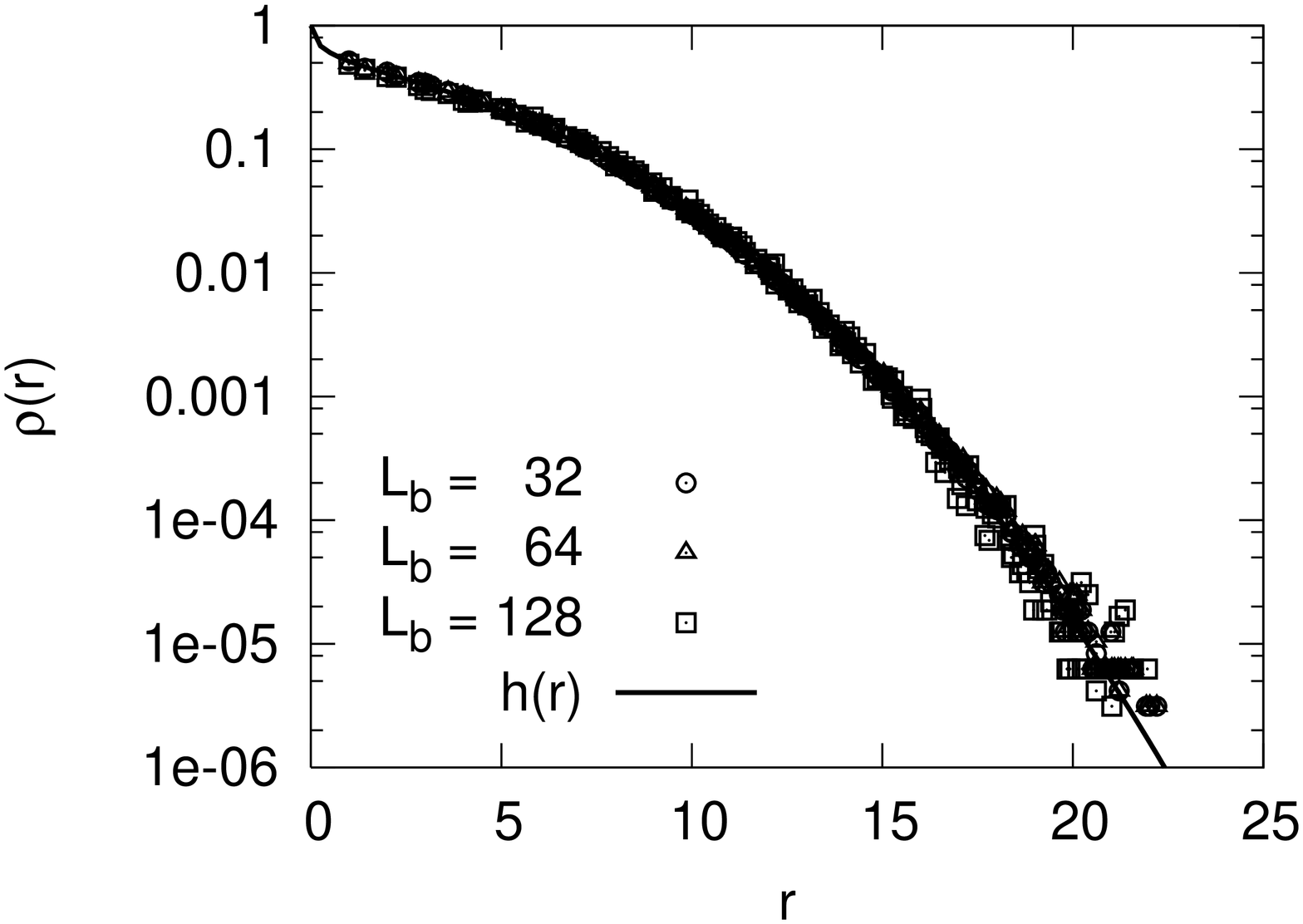} \\
\caption{(a) Radial distribution
function $\rho(r)$ plotted versus $r$ for side chain length
$N=50$, three choices of backbone length $L_b$ as indicated, and
the grafting density $\sigma =1$ for good solvent conditions.
 (b) Same as (a) but for $\Theta$-solvent conditions.
Parameters of the fit function $h(r)$
are quoted in the text.}
\label{fig12}
\end{center}
\end{figure}

When we look at the density profiles determined directly from simulations (see
Fig.~\ref{fig12} ), however, we recognize that there is no convex region in the interior of the
bottle brush, even at a grafting density of one which is the limit of what is
typically reached in experiment. Therefore, the comparison with worm-like
micelles is misleading, and we suggest to use an alternative form of fitting
function for the radial density
\begin{equation}\label{eq7a}
h(r) = \frac{\sigma}{1+(r/r_1)^{x_1}} \exp[-(r/r_2)^{x_2}]\; ,
\end{equation}
where $\sigma$ is the grafting density and $r_1$ and $r_2$ are the
length scales for the algebraic decay close to the backbone and the
exponential cutoff at larger distances (i.e., we expect $r_1 << r_2$
in the course of our fit analysis), and $x_1$ and $x_2$ are the
corresponding exponents. Taking into account the predictions of
scaling theory \cite{31} we can fix the first exponent $x_1 = (3\nu-1) /
2 \nu$. So again we are using three fit parameters. 
In Fig.~\ref{fig12} we show that this assumed form for the radial
density is able to fit the simulation data perfectly over almost six
orders of magnitude in density for both, good solvent and $\Theta$-solvent
conditions. The parameters of the shown fits are $r_1=0.49$,
$x_1=0.65$ ($\nu=0.588$), 
$r_2=10.67$, and $x_2=2.80$ for the good solvent case, and 
$r_1=1.19$, $x_1=0.5$ ($\nu = 0.5$), $r_2=7.13$,
and $x_2=2.18$ for the $\Theta$-solvent case.
Within the range of backbone lengths studied, the radial
density profiles agree, with some statistical fluctuations visible for
the good solvent data and the longest backbone, $L_b=128$.

\begin{figure}
\begin{center}
(a)\includegraphics[scale=0.29,angle=270]{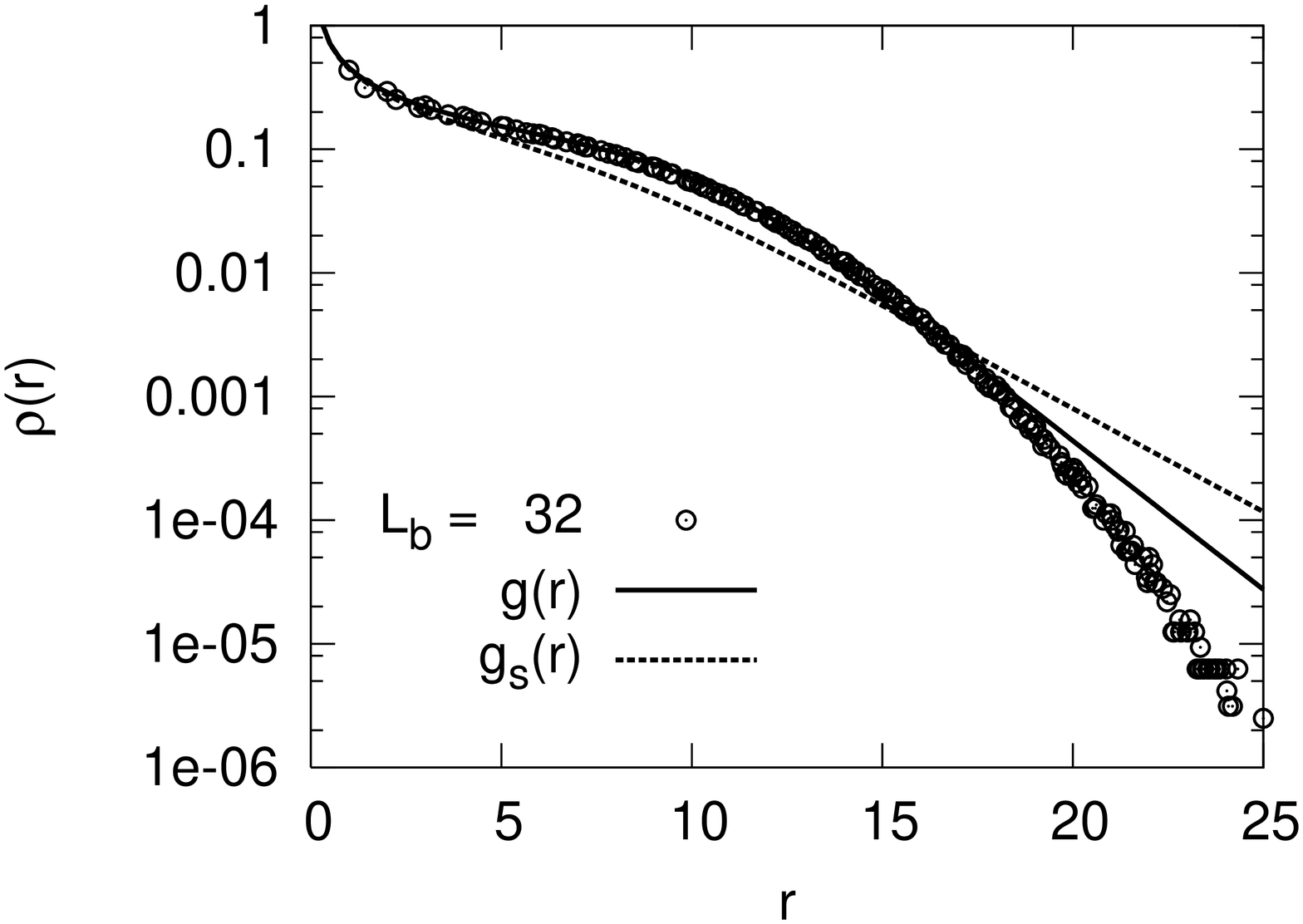}
\hspace{0.2cm}
(b)\includegraphics[scale=0.29,angle=270]{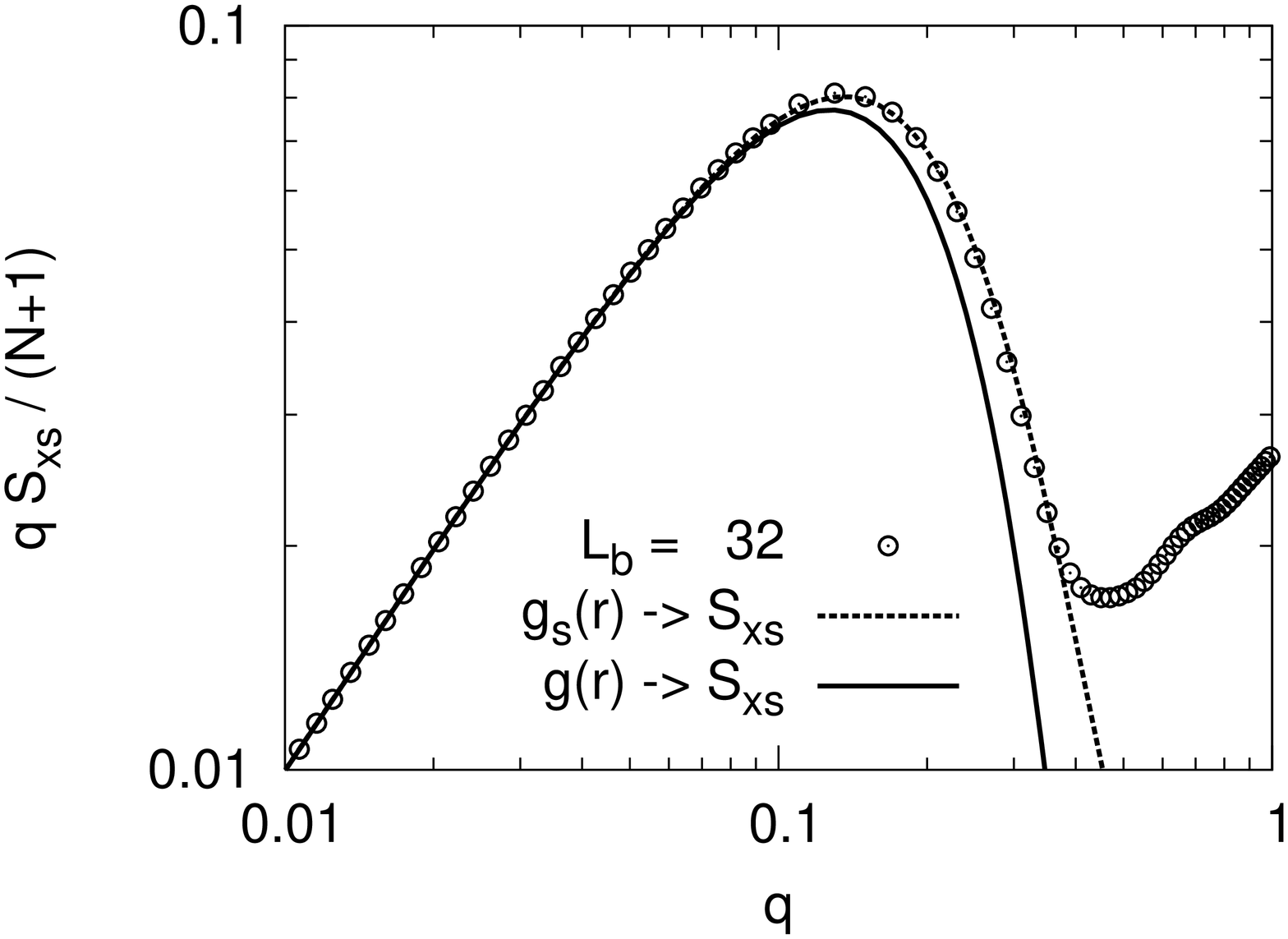} \\
\caption{(a) Radial distribution
function $\rho(r)$ plotted versus $r$ for side chain length
$N=50$ for good solvent conditions.
Parameters of the fit functions $g(r)$ (best fit to $\rho(r)$) and
$g_s(r)$ (Fourier transform of best fit to $S_{\rm xs}(q)$) 
are quoted in the text.
(b) The corresponding cross section structure
factor $S_{\rm xs}(q)=S_{\rm w}(q)/S_{\rm b}(q)$ plotted in the
representation $qS_{\rm xs}(q)$, vs. $q$.
The two curves correspond to the two curves in part (a).}
\label{fig13}
\end{center}
\end{figure}

\begin{figure}
\begin{center}
(a)\includegraphics[scale=0.29,angle=270]{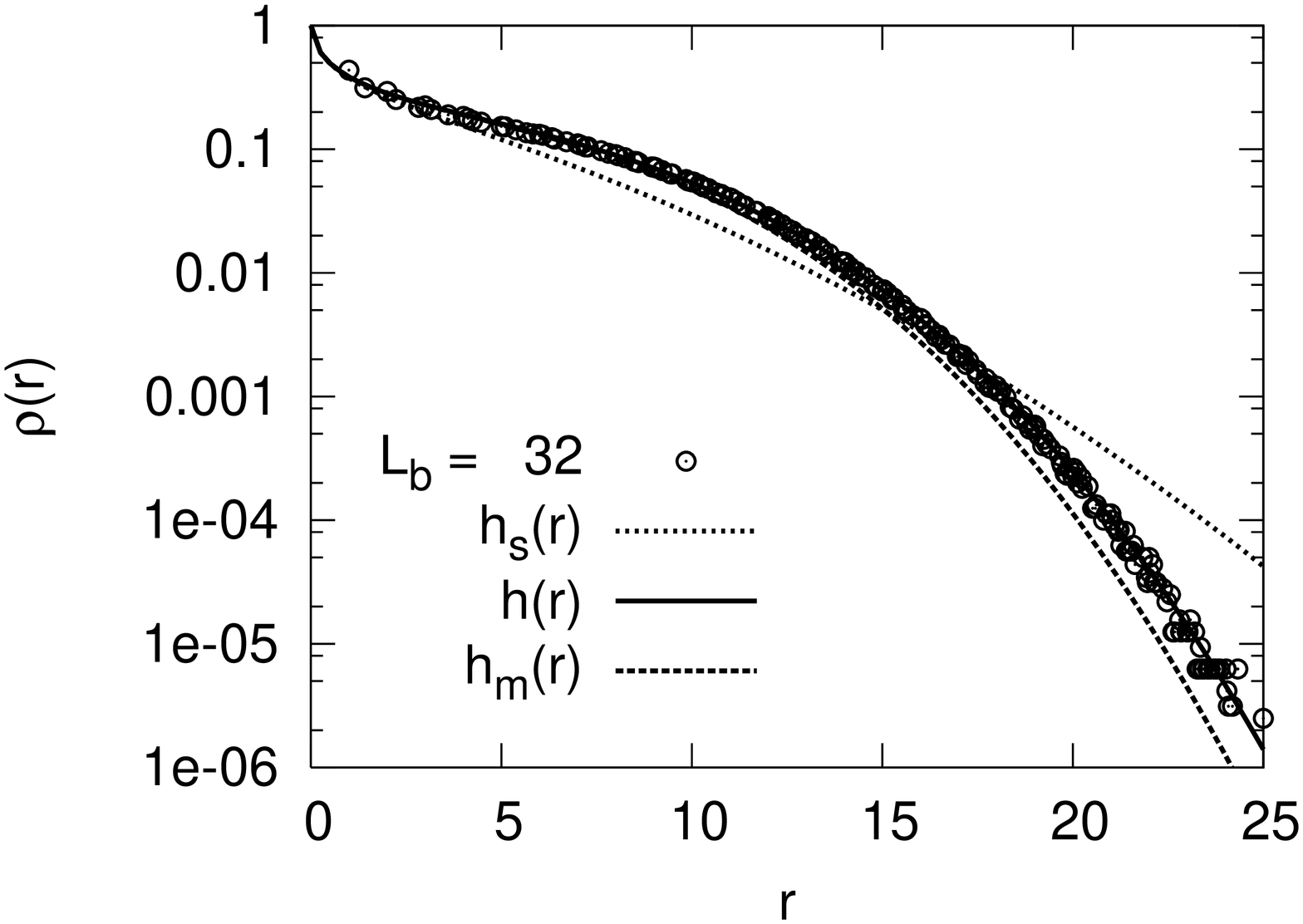}
\hspace{0.2cm}
(b)\includegraphics[scale=0.29,angle=270]{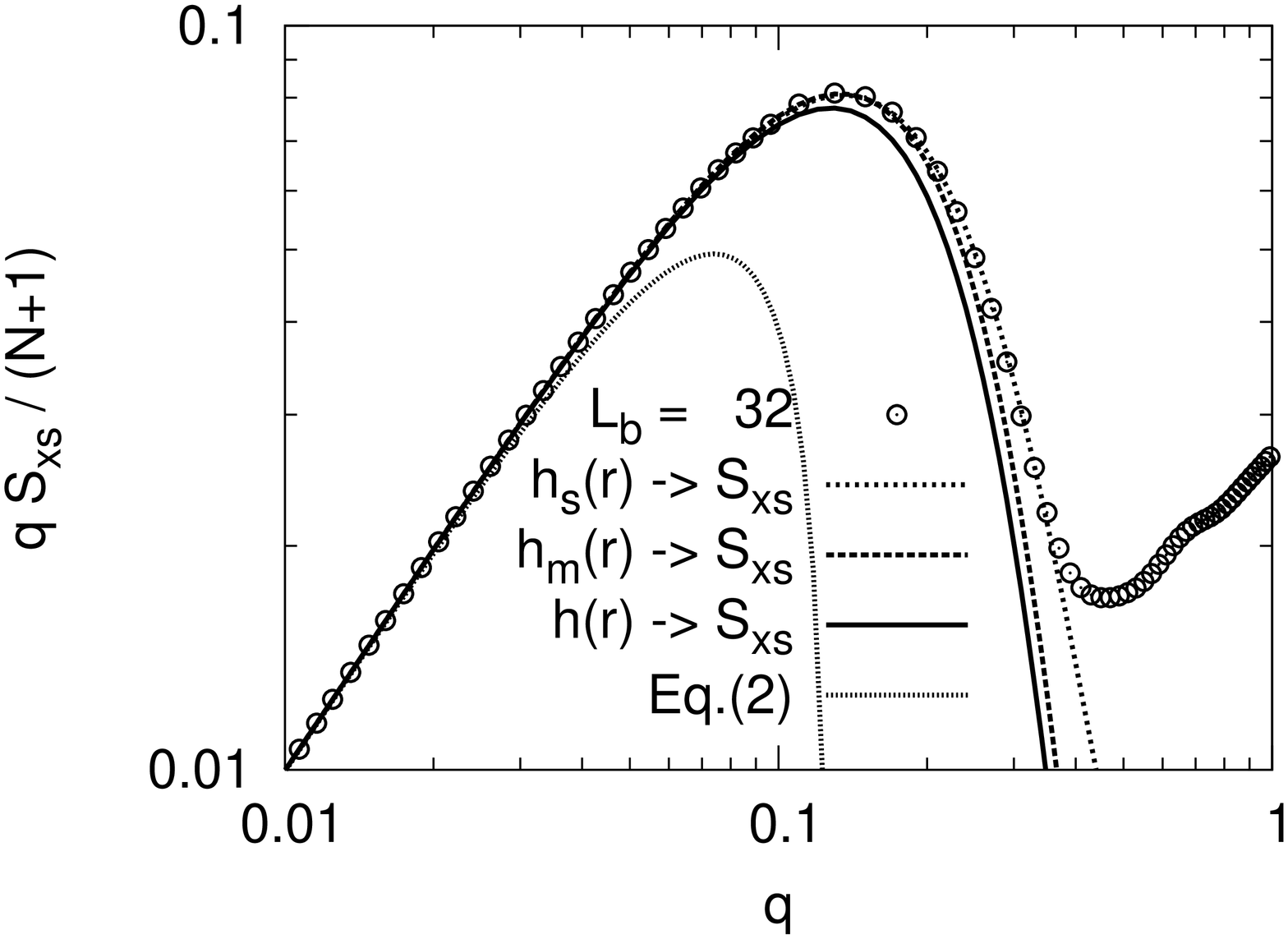} \\
\caption{Same as Fig.~\ref{fig13} but the fit functions are $h(r)$ and
  $h_s(r)$. Furthermore another fit is included which is extended up to
  the maximum in part (b) of this figure, $h_m(r)$.}
\label{fig14}
\end{center}
\end{figure}

Let us now turn to a discussion of the possibility to deduce the
radial density profile from the cross sectional structure factor as
defined in Eq.~(\ref{eq8}). Figs.~\ref{fig13}a and \ref{fig14}a show 
fits to the radial density profile using the functional forms $g(r)$
and $h(r)$ defined above. The form $g(r)$ suggested by Rathgeber el
al. \cite{2} is able to fit the radial density well over about $3$
orders of magnitude with parameters  $R_c=0.3$, $k=0.65$,
$R_s=10.5$ and $\sigma _s=1.90$; $h(r)$ fits over the complete range, as discussed
above. When we Fourier transform these functions according to
Eq.~(\ref{sxc-final}) and compare with the cross sectional structure factor
(full lines in Fig.~\ref{fig13}b and \ref{fig14}b), we
see that the transform only describes the scattering data well up to a
momentum transfer value of about $q=0.08$. This is only
slightly larger than the range over which one only sees the scattering from the large-scale
structure of the bottle brush (Eq.~(\ref{eq2})), which fits the cross
sectional structure factor up to a momentum transfer value of
$q\approx 0.04$, as shown by the dotted line
in Fig.~\ref{fig14}b. This regime then is basically determined
by the normalization of the radial density distribution. Using an
iterative optimization procedure \cite{numrecip} we can also find the best fit 
of the Fourier transform of the radial densities to the cross
sectional scattering shown by the curves indicated as $g_s(r)$ (fit
parameters are $R_c=0.3$, $k=0.65$,
$R_s=7.5$, and $\sigma_s=2.8$) and
$h_s(r)$ (fit parameters are $r_1=0.49$, $x_1=0.65$, $r_2=8.20$, and
$x_2=1.80$) in Figs.~\ref{fig13}b and ~\ref{fig14}b, where we extended the fit up
to $q\approx 0.4$. When we then look at these functions in real space
in Figs~\ref{fig13}a and \ref{fig14}a, we see that they are a rather poor fit to
the radial density profile. The function $h_m(r)$ 
(fit parameters are $r_1=0.49$, $x_1=0.65$, $r_2=10.20$, and $x_2=2.80$)
in Fig.~\ref{fig14}
will be discussed later in the text.

\begin{figure}
\begin{center}
(a)\includegraphics[scale=0.29,angle=270]{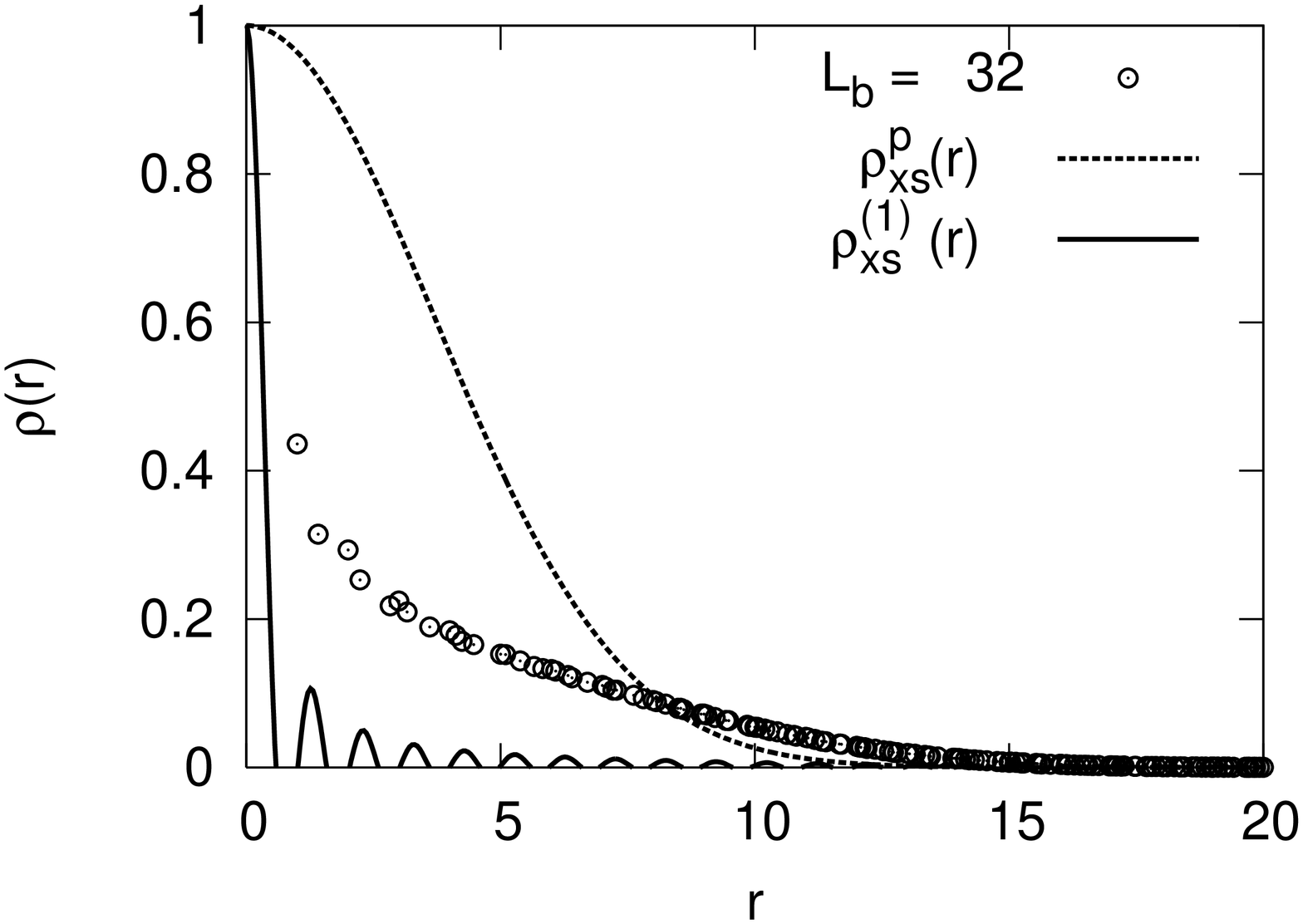}
\hspace{0.4cm}
(b)\includegraphics[scale=0.29,angle=270]{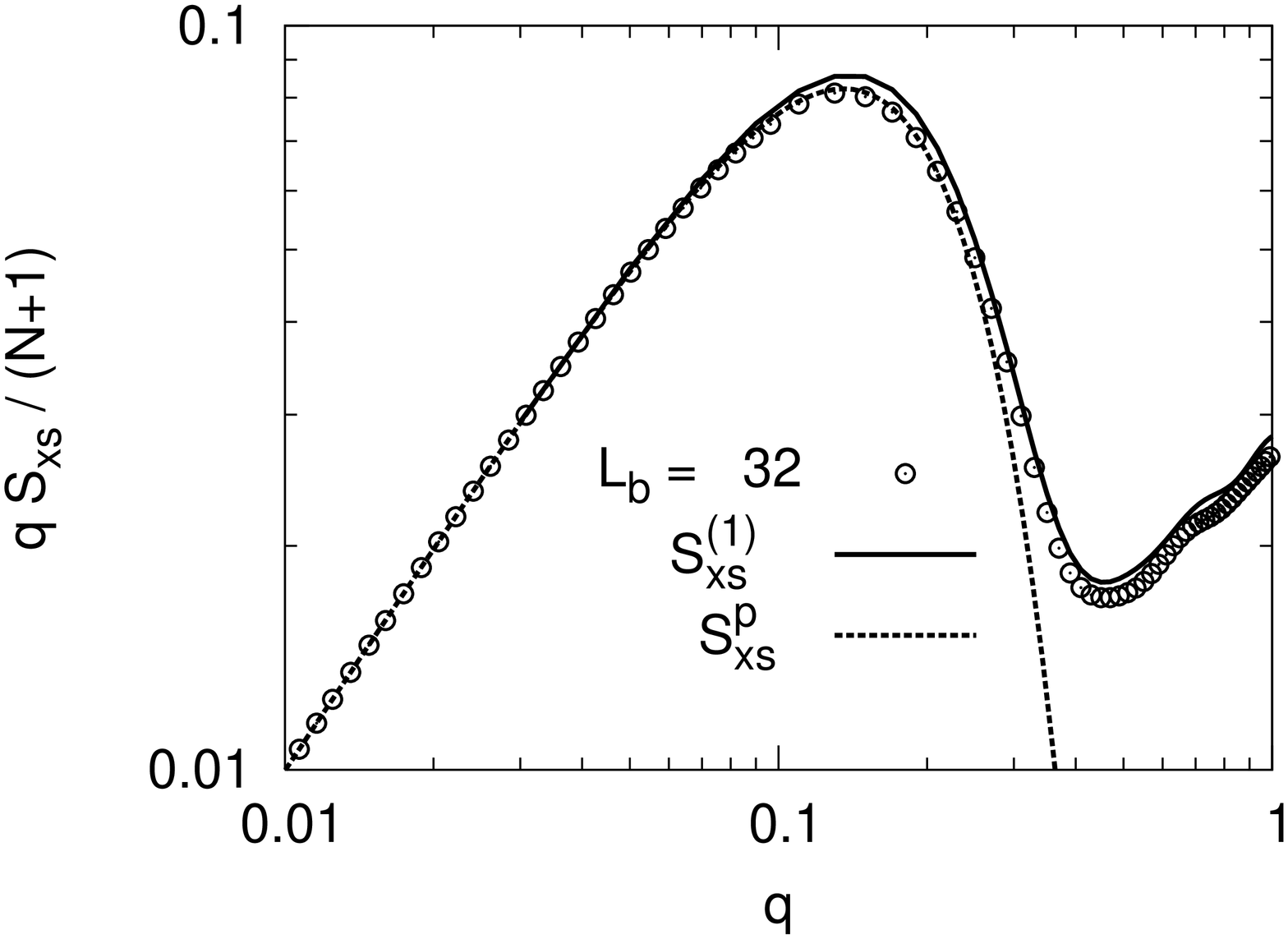}
\caption{(a) Monte Carlo data for the
radial distribution function $\rho(r)$ of the monomers plotted
versus $r$ for the side chain length $N=50$, $\sigma =1$ the good
solvent case, and backbone length $L_b=32$. The full curve labeled 
$\rho_{\rm xs}^{(1)}(r)$ shows the result of Bessel 
transforming (Eq.~(\ref{eq10})) the simulation data
for the scattering function into real space. 
$\rho_{\rm xs}^p(r)$ shows the prediction for $\rho(r)$ obtained
from fitting $s_{\rm xs}(q)$ using Eq.~(\ref{eq11})
(fit parameters are $R_c=1.0$, and $s=3.67$). (b)
The cross sectional scattering $qS_{\rm xs}(q)$ is plotted vs. $q$. Symbols
are data points, $S_{\rm xs}^{(1)}$ (full line) is the Bessel transform of the
full curve in part (a) which should ideally coincide with the symbols (see text).
$S_{\rm xs}^{p}$ (dashed line) is the best fit of Eq.~(\ref{eq11}) to the data.}
\label{fig15}
\end{center}
\end{figure}

\begin{figure}
\begin{center}
(a)\includegraphics[scale=0.29,angle=270]{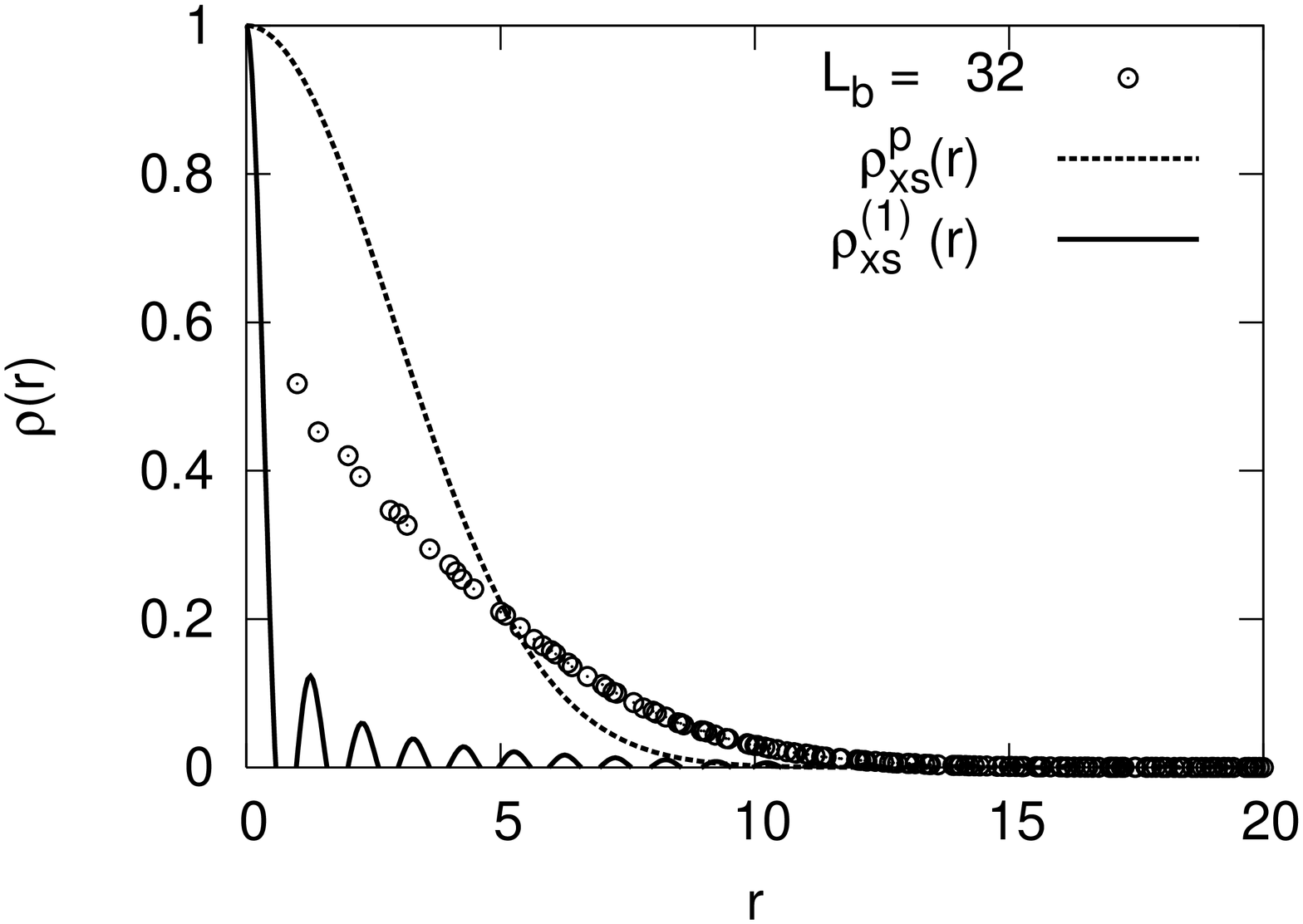}
\hspace{0.4cm}
(b)\includegraphics[scale=0.29,angle=270]{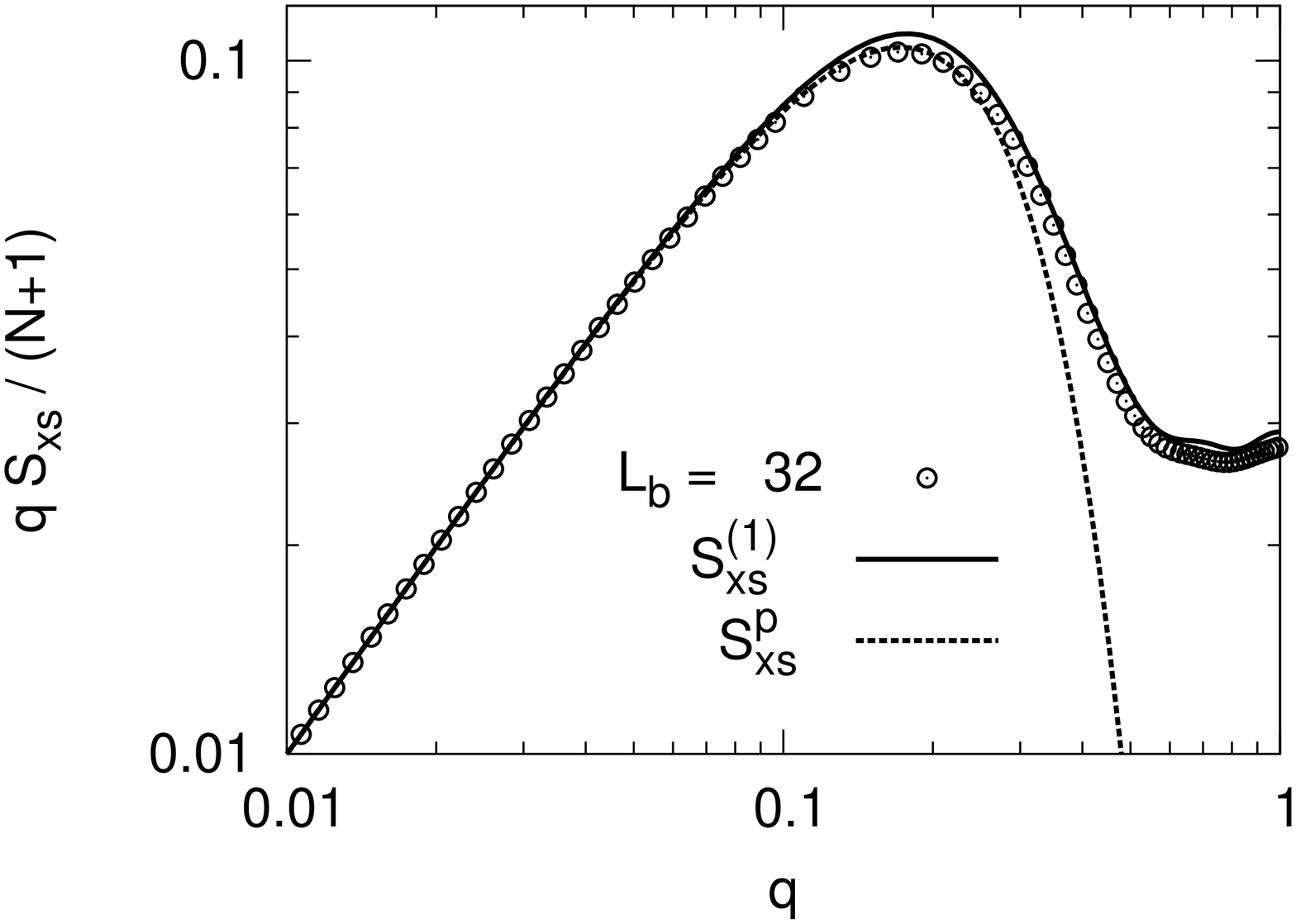}
\caption{Same as Fig.~\ref{fig15} for $\Theta$-solvent conditions.
(fit parameters are $R_c=1.0$, and $s=2.88$)}
\label{fig16}
\end{center}
\end{figure}

Using the functional form of Eq.~(\ref{eq11}) we can directly fit the
data in q-space and then transform into real space. This is shown in
Fig.~\ref{fig15} for the good solvent case and in Fig.~\ref{fig16} for the
$\Theta$-solvent case. Looking at Figs.~\ref{fig15}b and \ref{fig16}b one first
has to comment on the fact that the full curves in both figures
do not agree with the data given by the symbols. For these curves, the
scattering data in the q-range $[0,2\pi]$ were Bessel transformed into real
space and back again. The overestimation of the real scattering for
momentum transfers larger than about $0.1$ indicates that there is intensity
in the modes for q-values larger than $2\pi$ which is aliased into the studied
range. However, looking at the direct transform of the scattering data into
real space ($\rho_{\rm xs}^{(1)}$ in Figs.~\ref{fig15}(a) and \ref{fig16}(a)) one sees that
this is not a relevant numerical problem. Assuming the whole displayed q-range to be
relevant for the determination of the radial density profile leads to the
prediction of a highly oscillatory non-positive radial density. Similarly,
when we try to fit the scattering data beyond a q-value of about $0.4$ by the
assumed functional forms $g(r)$ and $h(r)$ we obtain unphysical radial density
profiles. Constraining the fit with the functional form of Eq.~(\ref{eq11}) to 
the q-range below $0.4$, however, also does not lead to a satisfactorily
prediction of the radial density as can be seen in Figs.~\ref{fig15}a and
\ref{fig16}a. The assumed convex shape of the radial density leads to an
overestimation of the density in the interior of the brush and a compensating
underestimation in the outer parts. 

Summarizing this discussion we have to conclude that there is only a small
range of momentum transfers where the analysis using
Eq.~(\ref{sxc-final}) may be warranted. This range extends at most to the position of the maximum in the
plot of $qS_{\rm xs}(q)$ vs. $q$. In this q-range, one should employ a concave
fitting function like the empirical law given by the function $h(r)$ above and
not the convex forms usually assumed for the inner part of the brushes. The
grafting densities typically employed in experiment are not high enough to lead
to a radial density which resembles a filled cylinder with a smeared out
interface to the solution when one works at good solvent or $\Theta$-solvent
conditions. This assumption may be valid working in poor solvent, a
regime which was not accessible to us using our simulation
approach. When we perform a fit to the cross sectional scattering only
for momentum transfers smaller than the peak position in the plots of
$q S_{\rm xs}(q)$ vs. $q$, we obtain the function $h_m(r)$ included in
Fig.~\ref{fig14}. We can see that this is a good representation
of the radial density down to values of about $\rho = 0.01$.

\begin{table}[htb]{\caption{Results for the cross sectional radius of
      gyration (see text) for the different fitting procedures and
      both solvent conditions.} \label{table1}}
\begin{tabular}{|l|lll|ll|l|}
\hline
  & $h(r)$ & $h_m(r)$ & $h_s(r)$ & $g(r)$ & $g_s(r)$ & $\rho_{\rm xs}^p(r)$\\
\hline
$R_{gc}$ (good solvent) & 7.83 & 7.49 & 7.64 & 7.94 & 7.79 & 5.24 \\ 
$R_{gc}$ ($\Theta$-solvent) & 6.19 & 5.87 & 5.98 & 6.39 & 6.27 & 4.08 \\
\hline
\end{tabular}
\end{table}
As a final result let us discuss the cross sectional radius of
gyration of the brush defined by
\begin{equation}
R_{gc}^2 = \frac{\int_0^\infty \rho(r) r^2 2\pi r
  dr}{\int_0^\infty \rho(r) 2\pi r dr} \; .
\end{equation}
Table 1 gives the resulting radii of gyration for the different
fitting functions and procedures employed and for the cases of good
solvent and $\Theta$-solvent. All fitting procedures reproduce the
shrinking of the brush going from good solvent to $\Theta$-solvent
condition. 
The results using the functional forms $h(r)$ and $g(r)$ agree
well with each other and also the suggested fit analysis of the 
scattering yielding function $h_m(r)$ results in only $4\%$
deviation from the true value. The fits using Eq.~(\ref{eq11}),
however, underestimate $R_{gc}$ by about $33\%$.

\section{CONCLUDING REMARKS}

In this paper, a comparative Monte Carlo study of bottle-brush
polymers with rigid and relatively long backbone lengths ($L_b=32$
to $L_b=256$ monomeric units) and flexible side chains of medium
length (up to $N=50$ monomeric units) under good solvent and 
$\Theta$-solvent conditions was performed, using the PERM algorithm. The
purpose of this study was to investigate the structure of such
macromolecules and to test physical assumptions used in experimental work on
related systems to extract structural information from scattering
data.

Our main results can be summarized as follows:

\begin{description}

\item{(i)} For the chosen side chain lengths, the chosen backbone
lengths already are clearly outside of the crossover regime from
bottle-brush to star polymer behavior. Comparing the total scattering function
$S_{\rm w}(q)$ of a bottle-brush polymer with and without pbcs
along the backbone, one does not find any
pronounced effect due to the different conformations the chains at the
end can assume in the two cases (therefore Fig.~\ref{fig9}
only shows the scattering for the free boundary case).
In addition, the range along
the backbone over which the effect of the proximity of the free
end of the backbone is felt in the side chain conformations is a
few monomer diameters only.

\item{(ii)}
Corroborating our earlier results~\cite{31} we find scaling
concepts in terms of power laws, blob pictures etc. not useful to
understand our results. We believe that scaling will become useful
if the chain lengths of the side chains are two orders of
magnitude larger; however, this limiting case is beyond the reach
of either experiment or simulation.

\item{(iii)} Correlations between backbone monomers and side chain
monomers do not contribute significantly to the scattering ,while
correlations between monomers from side chains anchoring at
different backbone positions do. As a consequence, the standard
factorization approximation by which the
cross-sectional scattering function $S_{\rm xs}(q)$ is related 
via Fourier transform to the radial monomer density
profile $\rho(r)$, is invalid for most of the momentum transfer range
typically studied. While experiments typically are
done for bottle-brush polymers with flexible backbones and we deal
here with the case of rigid backbones only, there is no reason why
approximations that are inaccurate in the latter case should
become accurate in the flexible backbone case, of course. From a
detailed analysis of the scattering function and radial density
obtained in the simulation we identify the regime where the analysis
of the cross sectional scattering might be successfully performed to
lie at q-values smaller than the position of the peak in the curves of
plots of $q S_{\rm xs}(q)$ vs. $q$. Here one should fit the Fourier transform of
a concave form of radial density dependence, as given, e.g., by Eq.(\ref{eq7a}).

\item{(iv)}
It would be desirable to perform neutron scattering from bottle
brushes where only a small fraction of side chains is deuterated.
In this way, a more direct information on the local conformational
structure in a bottle-brush could be gained, and more extensive
comparison with simulations should become possible. We also hope
that our study will stimulate further experimental work on bottle
brushes, in particular on the effect of solvent quality.
\end{description}

\underline{Acknowledgments}: One of us (H.-P. H.) received
financial support from the Deutsche Forschungsgemeinschaft (DFG)
via Sonderforschungsbereich SFB 625/A3. We are grateful to S.
Rathgeber and M. Schmidt for many stimulating discussions.

\end{document}